\documentclass[aps,prd,reprint,superscriptaddress,showpacs,nofootinbib]{revtex4-1}

\usepackage{hyperref}
\usepackage{amsmath}
\usepackage{amssymb}
\usepackage{graphicx}
\usepackage{color}
\usepackage{subfig}
\usepackage{bm}
\usepackage{gensymb}
\usepackage{bigints}
\usepackage[usenames,dvipsnames,svgnames,table]{xcolor}
\usepackage{pifont}
\usepackage{aas_macros}
\usepackage{orcidlink}
\usepackage{physics}

\usepackage[labelfont=bf]{caption}

\hypersetup{
    colorlinks=true,
    linkcolor=Blue,
    filecolor=Blue,
    urlcolor=MidnightBlue,
    citecolor=Blue
}

\numberwithin{equation}{section}

\begin{document}


\title{Metric-affine effects in crystallization processes of white dwarfs}

\author{Surajit Kalita\orcidlink{0000-0002-3818-6037}}
\email[E-mail: ]{surajit.kalita@uct.ac.za}
\affiliation{High Energy Physics, Cosmology \& Astrophysics Theory (HEPCAT) Group, Department of Mathematics \& Applied Mathematics, University of Cape Town, Cape Town 7700, South Africa}
\author{Lupamudra Sarmah\orcidlink{0000-0003-1651-9563}}
\email[E-mail: ]{lupamudra.sarmah@iiap.res.in}
\affiliation{Indian Institute of Astrophysics, Bengaluru 560034, India}
\affiliation{Department of Physics, Indian Institute of Technology (BHU), Varanasi 221005, India}
\author{Aneta Wojnar\orcidlink{0000-0002-1545-1483}}
\thanks{Corresponding author}
\email[E-mail: ]{awojnar@ucm.es}
\affiliation{Laboratory of Theoretical Physics, Institute of Physics, University of Tartu,
W. Ostwaldi 1, 50411 Tartu, Estonia
}
\affiliation{Departamento de F\'isica Te\'orica, Universidad Complutense de Madrid, E-28040, Madrid, Spain}

\begin{abstract}
We analyze the effects of modified gravity on specific heats of electrons and ions, Debye temperature, crystallization process, and cooling mechanism in white dwarfs. We derive the Lane-Emden-Chandrasekhar equation and relate it to the cooling process equations for Palatini $f(R)$ gravity. Moreover, for the first time in the literature, we show that the gravity model plays a crucial role not only in the mass and size of the white dwarf, but also affects their internal properties. We further demonstrate that modified gravity can decrease the cooling age significantly. 
\end{abstract}


\maketitle

\section{Introduction}
White dwarfs (WDs) are burned-out cores of main sequence stars with mass~$\lesssim (10\pm2)\, M_\odot$ and constitute the final stage of a stellar evolution~\citep{1986bhwd.book.....S,2018MNRAS.480.1547L}. The mass of a typical WD is similar to that of the Sun, but it is much smaller in size~(recent discovery shows a WD with the size of the moon~\cite{2021Natur.595...39C}). Due to its small radius, it generally has a low luminosity but very high surface temperature and thus occupies the lower left portion of the Hertzsprung-Russell (H-R) Diagram~\citep{2022AJ....164..126A}. As a star enters the WD phase, it should be noted that the only significant source of radiation comes from the residual ion thermal energy. Since there is no derivation of energy from the thermonuclear reactions at this phase of the stars, their ages significantly depend on the cooling processes. The thermal energy in the interior of a WD is much smaller than the Fermi energy and thus their structures are supported by the pressure of the degenerate electron gas. However, near the surface when the density drops significantly with respect to that of the core, thermal pressure starts becoming prominent. Many WDs are found in binary systems. When such a WD accretes matter from its companion, its mass increases. But there is a limit up to which this mass can increase and this limit is the well-known Chandrasekhar mass-limit (approximately $1.4\, M_\odot$ for a carbon-oxygen, non-rotating, non-magnetized WD)~\citep{1935MNRAS..95..207C}. Once a WD exceeds this critical mass, it undergoes a tremendous explosion in the form of a type Ia supernovae (SNe\,Ia). SNe\,Ia generally have consistent peak luminosities and thus are often used as standard candles~\citep{1997Sci...276.1378N,1987ApJ...323..140L}. However, the recent observations of several over-luminous~\citep{2006Natur.443..308H,2010ApJ...713.1073S,2009ApJ...707L.118Y,2013MNRAS.430.1030S} and under-luminous peculiar SNe\,Ia~\citep{1992AJ....104.1543F,1998AJ....116.2431T,2008MNRAS.385...75T} seem to question the uniqueness of the Chandrasekhar mass-limit as well as the existing theory of General Relativity (GR). It has been found that the over-luminous SNe\,Ia violate the Khokhlov pure detonation limit~\citep{1993A&A...270..223K} with a surprisingly large $^{56}$Ni mass content of $1.8\,M_\odot$ and thus one of the probable candidates for such SNe\,Ia are the super-Chandrasekhar WDs~\citep{2006Natur.443..308H}. Similarly, under-luminous SNe\,Ia with $^{56}$Ni mass as low as $0.05\,M_\odot$ infer indirect evidence of sub-Chandrasekhar limiting mass WDs.

These peculiar observations along with a few other astrophysical and cosmological problems motivated various groups around the world to modify Einstein's theory of GR. One such popular modification is the $f(R)$ gravity, with $R$ being the scalar curvature~\citep{1970MNRAS.150....1B,2011PhR...509..167C}. It aims at generalizing the Lagrangian density with an arbitrary function of $R$ and the modified field equations are obtained by following one of the two variational principles: the standard metric variation and the Palatini variation\footnote{Then, the curvature scalar $R$ is constructed with two independent objects, the metric and Ricci tensor built of the independent connection, see the details in Sec.~\ref{secpal}.}. In the metric approach, the action is varied with respect to the metric, leading to fourth-order field equations\footnote{It turns out, however, that the theory can be recast into the scalar-tensor representation from which it is easy to see that the $f(R)$ metric gravity possesses an extra degree of freedom, related to the scalar curvature~\cite{sotiriou2010f}.}. On the other hand, in the Palatini $f(R)$ gravity, the metric and connections are treated as independent of each other and the action is varied with respect to both of them~\citep{2007PhRvD..75f3509F,2006CQGra..23.1253S}. It must be noted that both these approaches, in general, give rise to entirely different theories and spacetime structures. This is due to the fact that the connection in the metric formalism is taken to be the Levi-Civita, a spacetime metric, whereas the connection in the Palatini formalism depends on the form of the chosen $f(R)$. Only when $f(R)$ is assumed to be a linear function of $R$, both these formalisms become equivalent and reduce to GR. In recent times, a number of phenomena starting from cosmic acceleration~\citep{2006PhRvD..73f3515S,2004GReGr..36.1765N,2006A&A...454..707A,2016JCAP...01..040B}, dark energy problem, compact objects including black holes, neutron stars, and WDs have extensively been studied using the Palatini $f(R)$ gravity~\citep{2017GReGr..49...25T,2021EPJC...81..888H,2019PhRvD.100d4020O}. 

So far, $f(R)$ gravity has been successful in explaining the peculiarities of WDs and this is evident by the large number of works being done in this regard. After initiating the exploration using metric $R+\alpha R^2$ gravity~\citep{2015JCAP...05..045D}, some of us later showed that the introduction of higher-order corrections to this model could get rid of the possible ghost modes as well as probe both the regimes of super- and sub-Chandrasekhar limiting mass WDs by just changing the central density~\citep{2018JCAP...09..007K}. Moreover, due to the comparatively large radius of WDs, Newtonian treatment is also viable and some works have been done by deriving the hydrostatic equilibrium equations in the weak-field limit for the metric~\citep{2022PhLB..82736942K,astashenok2022maximal} as well as the Palatini $f(R)$ gravity~\citep{2022PhRvD.105b4028S} along with the study of the stability of these modified gravity inspired WDs. Furthermore, by studying the gravitational wave signatures from these peculiar WDs, proposals were made to detect them by some futuristic gravitational wave detectors~\citep{2021ApJ...909...65K}.

A few tests with the use of stellar and sub-stellar objects have also been proposed. We know that the most fundamental equations to describe a (non-)relativistic star are the hydrostatic equilibrium equations along with the equation of state. Thus, altering any of them by the effects introduced by modified gravity results in the change of stars' internal properties (for a detailed review, see~\cite{Olmo:2019flu}), and thereby a different stellar evolution. They manifest by, for example, variations of the limited masses: the Chandrasekhar mass-limit of WDs, the minimum Main Sequence mass, minimum mass for deuterium burning, Jeans and opacity mass, etc. Seismic properties in stars~\cite{saltas2019obtaining} and terrestrial planets~\cite{Kozak:2021ghd,Kozak:2021zva,Kozak:2021fjy} turn out to be also affected, providing us tools to constrain theories. Another effect is observed in the light elements' abundances in stellar atmospheres~\cite{Wojnar:2020frr}. As mentioned, the evolutionary phases of non-relativistic stars, brown dwarfs, WDs~\cite{2022PhRvD.105b4028S}, and giant planets are also modified, and some of those phenomena, with the more accurate data provided by GAIA\footnote{\url{https://www.esa.int/Science_Exploration/Space_Science/Gaia/Gaia_overview}}, James Webb Space Telescope\footnote{\url{https://www.nasa.gov/mission_pages/webb/about/index.html}}, or Nancy Grace Roman Space Telescope\footnote{\url{https://www.nasa.gov/content/goddard/nancy-grace-roman-space-telescope}}, can also be used to constrain theories of gravity. For more discussion, see \cite{Wojnar:2022txk} and references therein.

In what follows, we are interested in the effects of gravitation on the processes happening in the interiors of WDs. Along with the violation of the well-known Chandrasekhar mass-limit, we also expect a modification in the physical processes of WDs. Therefore, we are interested in studying the crystallization process in the framework of Palatini $f(R)$ gravity. The cooling process of WDs can be roughly divided into four stages, namely, neutrino cooling, fluid cooling, crystallization, and Debye cooling~\citep{1986bhwd.book.....S}. Note that latent heat and sedimentation play a major role in crystallization. The net effect in this process is the migration of heavier elements towards the central region with a release of gravitational energy. Moreover, as the WD core crystallizes into a solid, latent heat is released which provides a source of thermal energy and hence delays the cooling process. Moreover, the crystallization temperature increases with the central density, and hence the massive WDs crystallize to solid at higher temperatures. Once the crystallization sets in and the star solidifies, the specific heat follows the Debye law and this is known as Debye cooling. Because Palatini $f(R)$ gravity modifies the hydrostatic equilibrium equations and the resulting mass--radius curve, it does modify the other physical properties such as specific heat, Debye temperature, and hence the crystallization process, as we demonstrate in this paper.

Following the indications discussed in the previous works on how the gravitational interaction affects microphysical properties\footnote{Such as a gravitational dependence of the chemical potential and temperature~\cite{kulikov1995low,li2022we}, equations of state~\cite{kim2014physics,Wojnar:2022dvo}, opacities~\cite{sakstein2015testing}, energy generation rates~\cite{sakstein2015hydrogen,Olmo:2019qsj,rosyadi2019brown,Wojnar:2020frr}, chemical reactions rates~\cite{lecca2021effects}, and elementary particle interactions~\cite{delhom2018observable}.}, in this work, we revise the crystallization process~\cite{mestel1967energy,van1968crystallization} of WDs in modified gravity. As a simple generalization of Einstein's gravity, we base our calculations on Palatini $f(R)$ gravity, which we briefly recall in Sec.~\ref{secpal}. We further discuss the non-relativistic limit of this theory, modified hydrostatic equilibrium equation, and Chandrasekhar equation of state~\cite{1935MNRAS..95..207C}, which combined together, provides the so-called Lane-Emden-Chandrasekhar (LEC) equation for Palatini theory. In Sec.~\ref{secool}, we demonstrate that the Debye temperatures and specific heats do depend on the model of gravity, and as a consequence, it further affects the latent heat released during the crystallization process. Taking this correction into account, we also provide a more accurate cooling model of WDs. Finally, Sec.~\ref{secon} includes a discussion of these results and our conclusions.


\section{Palatini $f(R)$ gravity formalism and modified hydrostatic balance equations}\label{secpal}

Let us first briefly discuss the basic formalisms of Palatini $f(R)$ gravity. Throughout this paper, we consider the metric signature to be $(-,+,+,+)$. For a spacetime metric $g_{\mu\nu}$, the action for $f(R)$ gravity is given by~\cite{2010LRR....13....3D}
\begin{align}
    S[g,\Gamma,\Psi]=\frac{1}{2\kappa}\int\sqrt{-g}f\big(R(g,\Gamma)\big)\dd[4]{x} + S_\mathrm{m}[g,\Psi],
\end{align}
where $\kappa=-8\pi G/c^4$, $g=\det(g_{\mu\nu})$, $G$ is Newton's gravitational constant, $c$ is the speed of light, and $S_\mathrm{m}$ is the matter action which depends on the metric and the matter field $\Psi$. Varying this action with respect to $g_{\mu\nu}$ results in the modified field equation, given by~\citep{2010LRR....13....3D}
\begin{equation}\label{Eq: f(R) master eq}
f'(R)R_{\mu\nu}-\frac{1}{2}f(R)g_{\mu\nu}=\kappa \mathcal{T}_{\mu\nu},
\end{equation}
where $f'(R) = \dv*{f(R)}{R}$ and $\mathcal{T}_{\mu\nu}$ is the energy-momentum tensor. It is easy to verify that in GR, $f(R)=R$, and thus this equation reduces to the famous Einstein equation. Moreover, varying $S$ with respect to $\Gamma$ gives the following modified equation:
\begin{equation}
\nabla_{\lambda}\left(\sqrt{-g}f'(R)g^{\mu\nu}\right)=0,
\end{equation}
where $\nabla_{\lambda}$ is the covariant derivative ruled by $\Gamma$. Let us now define a new metric tensor $\bar g_{\mu\nu}$ such that $\bar g_{\mu\nu}= f'(R)g_{\mu\nu}$. Thus the above equation simplifies to
\begin{equation}
\nabla_\lambda(\sqrt{-\bar g}\bar g^{\mu\nu})=0,
\end{equation}
providing now that $\Gamma$ is the Levi-Civita connection of the metric $\bar g_{\mu\nu}$.

Assuming the matter of the WD is non-magnetized and behaves like a perfect fluid, the energy-momentum tensor can be written as
\begin{equation}
    \mathcal{T}^{\mu\nu} = \left(\rho c^2 + P\right)u^{\mu}u^\nu + Pg^{\mu\nu},
\end{equation}
where $P$ and $\rho$ are respectively the pressure and matter density of the fluid. Assuming the Newtonian limit, we have $\rho c^2\gg P$. Now, in the weak-gravity limit, expanding the metric tensors as $g_{\mu\nu} = \eta_{\mu\nu}+h_{\mu\nu}$ and $\bar{g}_{\mu\nu}= \eta_{\mu\nu}+\bar{h}_{\mu\nu}$, such that $\abs{h_{\mu\nu}}$, $\abs{\bar{h}_{\mu\nu}}\ll\abs{\eta_{\mu\nu}}$, and substituting it in Eq.~\eqref{Eq: f(R) master eq} with an analytic functional
\begin{equation}
    f(R)=\sum_{i=0}\alpha_i{R}^i,
\end{equation}
one obtains the following modified Poisson equation~\cite{toniato2020palatini}
\begin{equation}
    \nabla^2\Phi\approx 4\pi G \left(\rho - 2\alpha\nabla^2\rho\right),
\end{equation}
where $h_{00}=-2\Phi/c^2$ with $\Phi$ being the gravitational potential. Note that one may neglect the contribution from cosmological constant in stellar studies \cite{liu2019properties} and thus $\alpha_0=0$. Moreover, without the loss of generality, one can safely assume $\alpha_1=1$ and thus in the above equation, $\alpha$ comes from the quadratic term i.e., $\alpha:=\alpha_2$. Therefore, the non-relativistic objects feel contributions from the linear and quadratic terms of the gravitational Lagrangian, i.e. effectively we have $f(R)= R+\alpha R^2$.

In the weak-field limit, for a spherically symmetric spacetime, the hydrostatic-balance and the mass-estimate equations can be respectively written as
\begin{align}
    \dv{\Phi}{r} &= -\frac{1}{\rho}\dv{P}{r}, \\
    \dv{m}{r} &= 4\pi r^2 \rho.
\end{align}
Using these relations in the modified Poisson equation, we obtain
\begin{equation}\label{poisson}
    \frac{1}{r^2}\dv{r}(\frac{r^2}{\rho}\dv{P}{r}) = -4\pi G\left[\rho-\frac{2\alpha}{r^2}\dv{r}(r^2\dv{\rho}{r})\right],
\end{equation}
which, upon simplification, can be recast as
\begin{equation}
    \dv{P}{r}= -\frac{Gm\rho}{r^2} + 8\pi G \alpha \rho \dv{\rho}{r}.
\end{equation}

Since the interiors of WDs predominantly consist of electron-degenerate matter, we consider the Chandrasekhar equation of state (EoS) to obtain their structures. Defining $p_\text{F}$ as the Fermi momentum, $m_\text{e}$ the mass of an electron, $h$ the Planck's constant, $\mu_\text{e}$ the mean molecular weight per electron, and $m_\text{p}$ the mass of a proton, the Chandrasekhar EoS is given by~\cite{1935MNRAS..95..207C}
\begin{equation}\label{Chandrasekhar EoS}
\begin{aligned}
    P &= \frac{\pi m_\text{e}^4 c^5}{3 h^3}\left[x_\text{F}\left(2x_\text{F}^2-3\right)\sqrt{x_\text{F}^2+1}+3\sinh^{-1}x_\text{F}\right],\\
    \rho &= \frac{8\pi \mu_\text{e} m_\text{p}(m_\text{e}c)^3}{3h^3}x_\text{F}^3,
\end{aligned}
\end{equation}
where $x_\text{F} = p_\text{F}/m_\text{e}c$. Denoting $A = \pi m_\text{e}^4 c^5/3 h^3$, $B = 8\pi \mu_\text{e} m_\text{p}(m_\text{e}c)^3/3h^3$, and $x_\text{F} = x$, this EoS can be written as
\begin{equation}
    P=Ag(x),\quad\rho = B x^3,
\end{equation}
where $g(x) = x\left(2x^2-3\right)\sqrt{x^2+1}+3\sinh^{-1}x$. Using this EoS, the modified Poisson equation~\eqref{poisson} can be recast as
\begin{equation}\label{Poisson2}
    \resizebox{1.0\hsize}{!}{$\frac{1}{r^2}\dv{r}(r^2\dv{\sqrt{x^2+1}}{r}) = -\frac{\pi GB^2}{2A}\left[x^3 - \frac{2\alpha}{r^2}\dv{r}(3r^2x^2\dv{x}{r})\right].$}
\end{equation}
Assuming $y^2=x^2+1$ such that at $r=0$, we have $x(0)=x_0$ and $y(0)=y_0$. We now introduce new variables $\eta$ and $\phi$, given by
\begin{equation}
    r=a\eta, \quad y=y_0\phi,
\end{equation}
with
\begin{equation}
    a=\left(\frac{2A}{\pi G} \right)^{1/2}\frac{1}{By_0},\quad
    y^2_0=x^2_0+1.
\end{equation}
Substituting these variables in the modified Poisson equation~\eqref{Poisson2}, we obtain the following modified Lane-Emden-Chandrasekhar~(LEC) equation for the Palatini $f(R)$ gravity
\begin{equation}\label{LEC}
    \resizebox{1.0\hsize}{!}{$\frac{1}{\eta^2} \dv{\eta}\left[\eta^2\dv{\phi}{\eta}\left\{1 - \frac{6\alpha}{a^2} \phi\left(\phi^2-\frac{1}{y^2_0}\right)^{1/2} \right\}\right] = -\left(\phi^2-\frac{1}{y^2_0}\right)^{3/2}.$}
\end{equation}
The boundary conditions at the center of the WD are $\phi(\eta=0)=1$ and $\phi'(\eta=0)=0$. The vanishing density at the surface provides $\phi(\eta_1)=1/y_0$, where the radius of the star $\mathcal{R}$ is given by $\mathcal{R} = a\eta_1$. Furthermore, from our definitions, the matter density is given by
\begin{equation}\label{density}
    \rho=\rho_\mathrm{c}\frac{y^3_0}{\left(y^2_0-1\right)^{3/2}}\left(\phi^2-\frac{1}{y^2_0}\right)^{3/2},
\end{equation}
where the central density $\rho_\mathrm{c}$ follows
\begin{equation}
    \rho_\mathrm{c}=Bx^3_0=B\left(y^2_0-1\right)^{3/2}.
\end{equation} 
Therefore, the total mass of the WD written with respect to the solution of the modified LEC equation is given by
\begin{align}
    \resizebox{1.0\hsize}{!}{$\mathcal{M} = -\frac{4\pi}{B^2}\left(\frac{2A}{\pi G} \right)^{3/2} \left[\eta^2\dv{\phi}{\eta} \left\{1-\frac{6\alpha}{a^2}\phi \left(\phi^2-\frac{1}{y^2_0}\right)^{1/2}\right\}\right]_{\eta=\eta_1}.$}
\end{align}
Note that at the surface, we already have $\phi(\eta_1)=1/y_0$, and thus the above expression reduces to 
\begin{equation}
   \left. \mathcal{M} = -\frac{4\pi}{B^2} \left(\frac{2A}{\pi G} \right)^{3/2} \left(\eta^2\dv{\phi}{\eta}\right)\right|_{\eta=\eta_1}.
\end{equation}

\begin{figure}[t]
	\centering
	\includegraphics[scale=0.4]{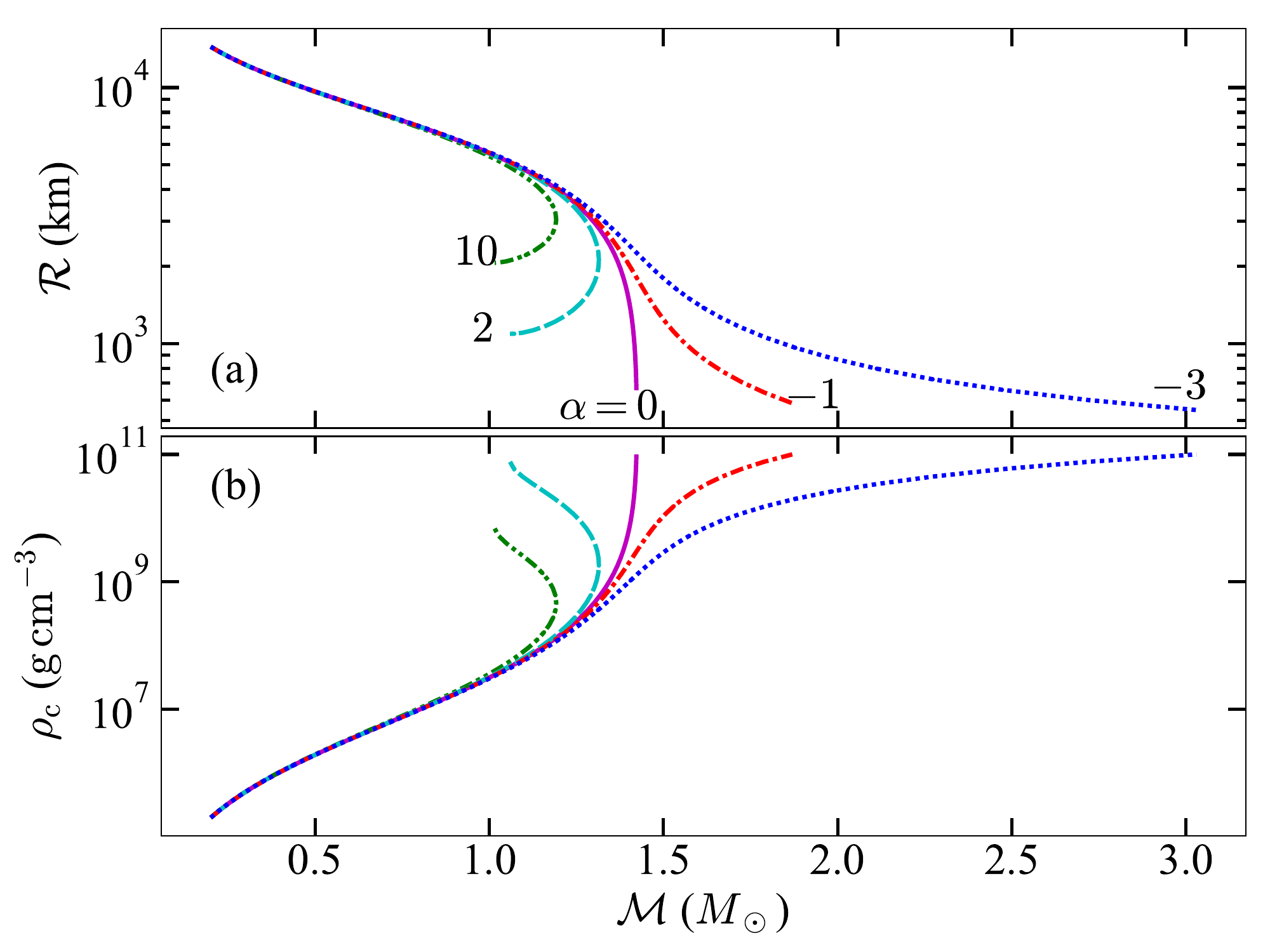}
	\caption{(a) Mass--radius relation and (b) central density as a function of the WD mass in Palatini $f(R)$ gravity. The numbers adjacent to the various lines denote $\alpha$ in the units of $10^{13} \rm\, cm^2$.}
	\label{Fig: MR}
\end{figure}
Before going further, let us discuss the mass--radius relation and constraints on the parameter $\alpha$, which will be useful for our further considerations. Fig.~\ref{Fig: MR} shows the mass--radius relation as well as the variation of mass with respect to $\rho_\mathrm{c}$ for the modified gravity inspired WDs. $\alpha=0$ corresponds to the Chandrasekhar mass--radius relation with a mass-limit of approximately $1.4\,M_\odot$. For $\alpha\neq0$, the mass--radius overlaps with the Chandrasekhar one at low density. For negative $\alpha$, it deviates in such a way that mass of the WD goes over this mass-limit and it can explain the origin of super-Chandrasekhar WDs resulting in the formation of over-luminous SNe\,Ia. On the other hand, for positive $\alpha$, it deviates on the other side after attaining a maximum mass. We already showed that in Palatini $f(R)$ gravity, decreasing mass with the increase in $\rho_\mathrm{c}$ results in unstable WDs, which can blow up under small radial perturbation~\cite{2022PhRvD.105b4028S}. Thus, the maximum mass for positive $\alpha$ is the mass-limit of the WD, which turns out to sub-Chandrasekhar and it can explain the origins of under-luminous SNe\,Ia. Therefore, this model of gravity can explain the sub- and super-Chandrasekhar limiting mass WDs depending on the values of $\alpha$. In fact, a better model would be the one where the model parameters are fixed throughout and both the mass regimes are obtained just by varying the central density. In other words, both the mass regimes can be explained by a single mass--radius curve. Such exploration was done previously for metric $f(R)$ gravity in~\cite{2018JCAP...09..007K,2022PhLB..82736942K}. However, the calculations are more complicated, and because in this paper, our target is to explain the thermal properties of the WDs, we use this simplified model.

Regarding the constraints, there are a few works to mention. In this work, we use the bound given by the Gravity Probe B experiment, which states that $|\alpha|\lesssim 5\times10^{15}\rm\,cm^2$~\cite{2010PhRvD..81j4003N}. Notice that in the case of Palatini gravity, the value of the model parameter is related to the curvature regime ~\cite{olmo2005gravity}, and because the Palatini curvature scalar is proportional to the density, its value also does. The analytical studies in the weak-field limit provided $|\alpha| \lesssim 2\times 10^{12}\rm\,cm^2$~\cite{olmo2005gravity}, while when electric forces taken into account and assumed to be of the same order of magnitude, resulted in $|\alpha| \lesssim 2\times 10^{9}\rm\,cm^2$~\cite{avelino2012eddington,jimenez2018born}. Further considerations revealed that the Solar System experiments do not deliver bounds on the parameters because of the microphysics uncertainties~\cite{toniato2020palatini}. Moreover, Palatini gravity, in a similar fashion as GR, is not able to explain the galaxy rotation curves~\cite{Hernandez-Arboleda:2022rim}. Thus the parameter in Palatini gravity has not been yet restrictively constrained, and hence, for a given $\rho_\mathrm{c}$, one deals with a particular range of the parameter's value.

\section{Cooling model of white dwarfs in Palatini $f(R)$ gravity}\label{secool}

In our previous work~\cite{Kalita:2022zki}, we studied a simplified cooling model in Palatini $f(R)$ gravity. In that approach, we assumed that the total thermal energy of the star ($3/2 k_\mathrm{B}T$ is the thermal energy per ion) is given by
\begin{equation} \label{Uthermal}
    U=\frac{3}{2} k_\mathrm{B}T\frac{\mathcal{M}}{A m_\mathrm{p}},
\end{equation}
where $T$ is the temperature of the isothermal core and $\mathcal{M}/(Am_\mathrm{p})$ is the number of ions with $A$ being the mean atomic weight.  It means that the main contribution comes from the ions (for which the specific heat $c_v=(3/2)k_\mathrm{B}$ when crystallization is not taken into account). In these calculations, we neglected its dependency on the Debye temperature $\Theta_\mathrm{D}$ and the ratio of Coulomb to thermal energy. In what follows, we are going to take into account those properties in this work. We eventually observe how modified gravity affects them and thereby the cooling process of WDs.

\subsection{Thermal heat of white dwarfs}

We now want to take into account the thermal properties of WDs' interior. Denoting the specific heats of ions as $c_v^\text{ion}$ and the same for the electrons per ions as $c_v^\text{el}$, the mean specific heat is given by
\begin{equation}\label{mean}
    \bar{c}_v=\frac{1}{\mathcal{M}}\int_0^\mathcal{M}(c_v^\text{el}+c_v^\text{ion})\dd{m}.
\end{equation}
This average is taken for the whole stellar configuration because both specific heats depend on density. Thus, the thermal energy of our star in that case is given by
\begin{equation}
    U=\bar{c}_v\frac{\mathcal{M}}{A m_\mathrm{p}} T,
\end{equation}
 and thereby the luminosity provided by the rate of decrease in thermal energy of ions and electrons in time $t$, takes the form
\begin{equation}\label{LU}
    L=-\dv{U}{t}=-\frac{\mathcal{M}}{A m_\mathrm{p}}\bar{c}_v\dv{T}{t}.
\end{equation}

Let us now discuss each element of Eq.~\eqref{mean}. The specific heat of the electrons per ion is given by~\cite{koester1972outer}
\begin{equation}
    c_v^\text{el}=\frac{3}{2}\frac{k_\mathrm{B}\pi^2}{3}Z \frac{k_\mathrm{B}T}{\epsilon_\mathrm{F}},
\end{equation}
where $Z$ is the charge and $\epsilon_\mathrm{F}$ is the Fermi energy, which is related to the Fermi momentum $p_\mathrm{F}$ as
\begin{align}\label{Eq: relativistic EF-pF}
\epsilon_\mathrm{F}^2 = p_\mathrm{F}^2c^2 + m_\mathrm{e}^2c^4,\\
p_\mathrm{F}^3 = \frac{3h^3}{8\pi}\frac{\rho}{\mu_\mathrm{e}m_\mathrm{p}}.
\end{align}

\begin{figure}[htpb]
	\centering
	\includegraphics[scale=0.5]{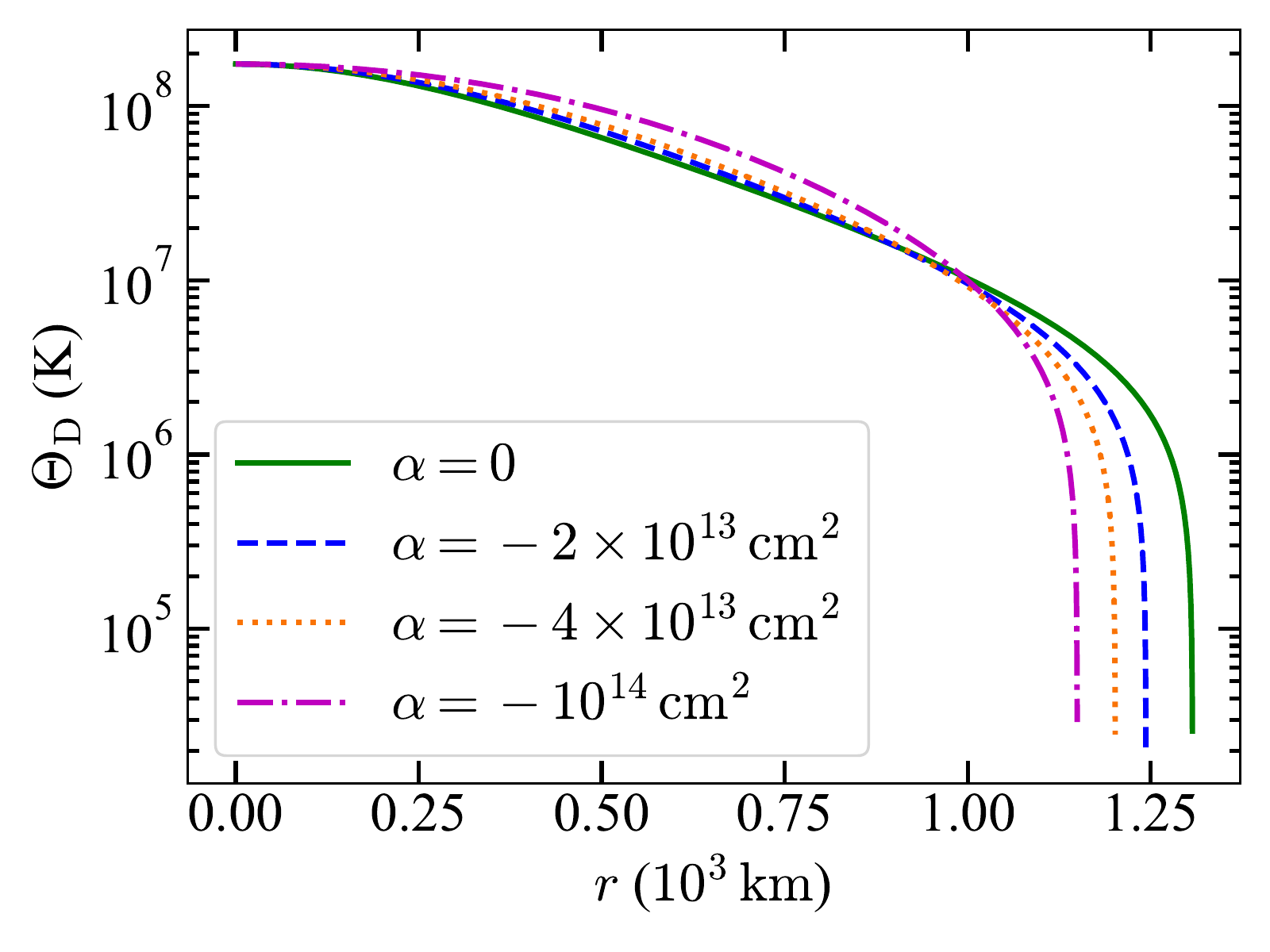}
	\caption{Variation of the Debye temperature inside a modified gravity inspired carbon WD with $\rho_\mathrm{c}=10^{10}\rm\,g\,cm^{-3}$.}
	\label{debtemp}
\end{figure}

\begin{figure}[htbp]
     \subfloat[$T = 10^6$\,K]{%
     \centering
       \includegraphics[scale = 0.42]{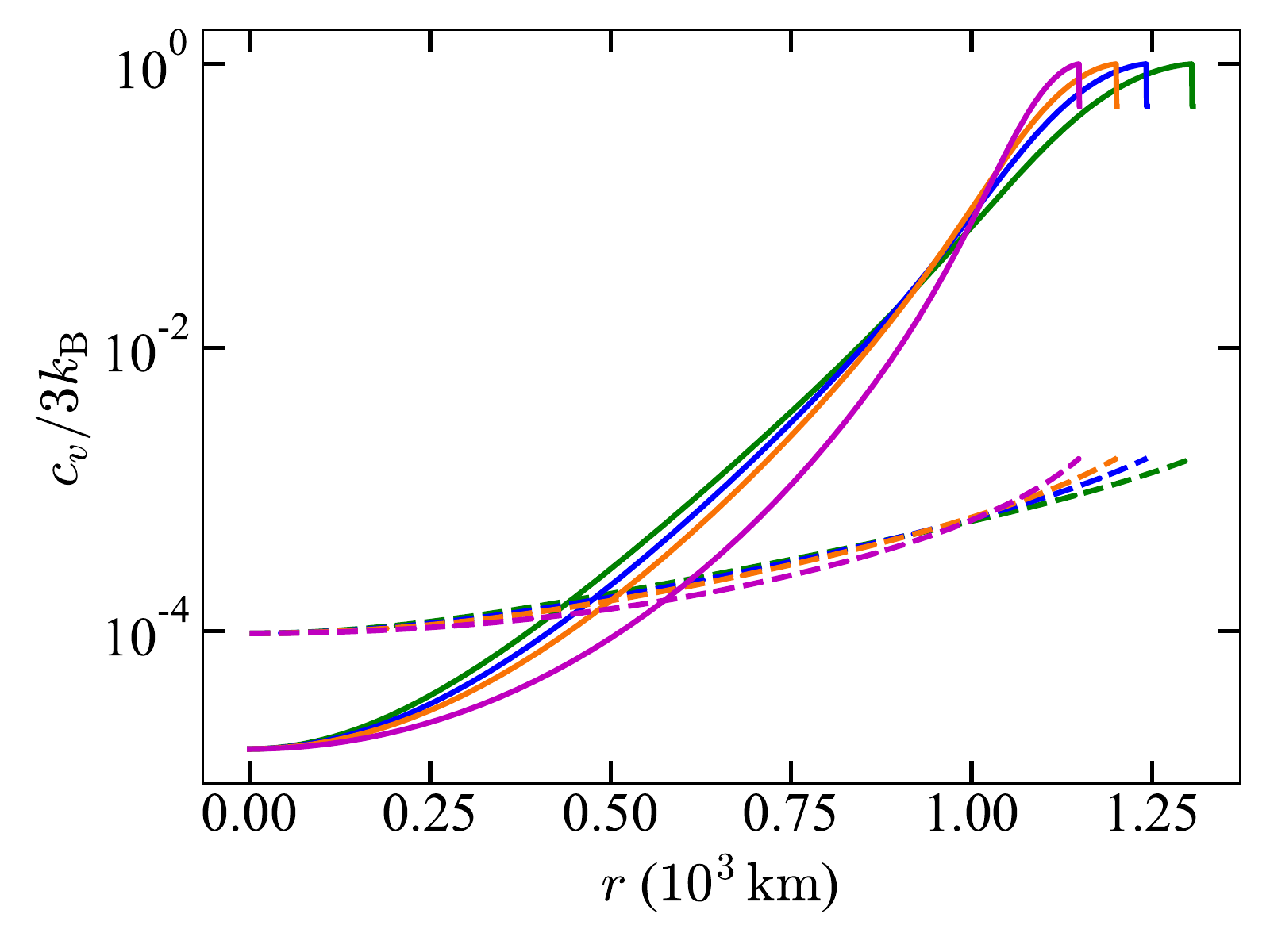}
    }\\
     \subfloat[$T = 10^7$\,K]{%
     \centering
       \includegraphics[scale = 0.42]{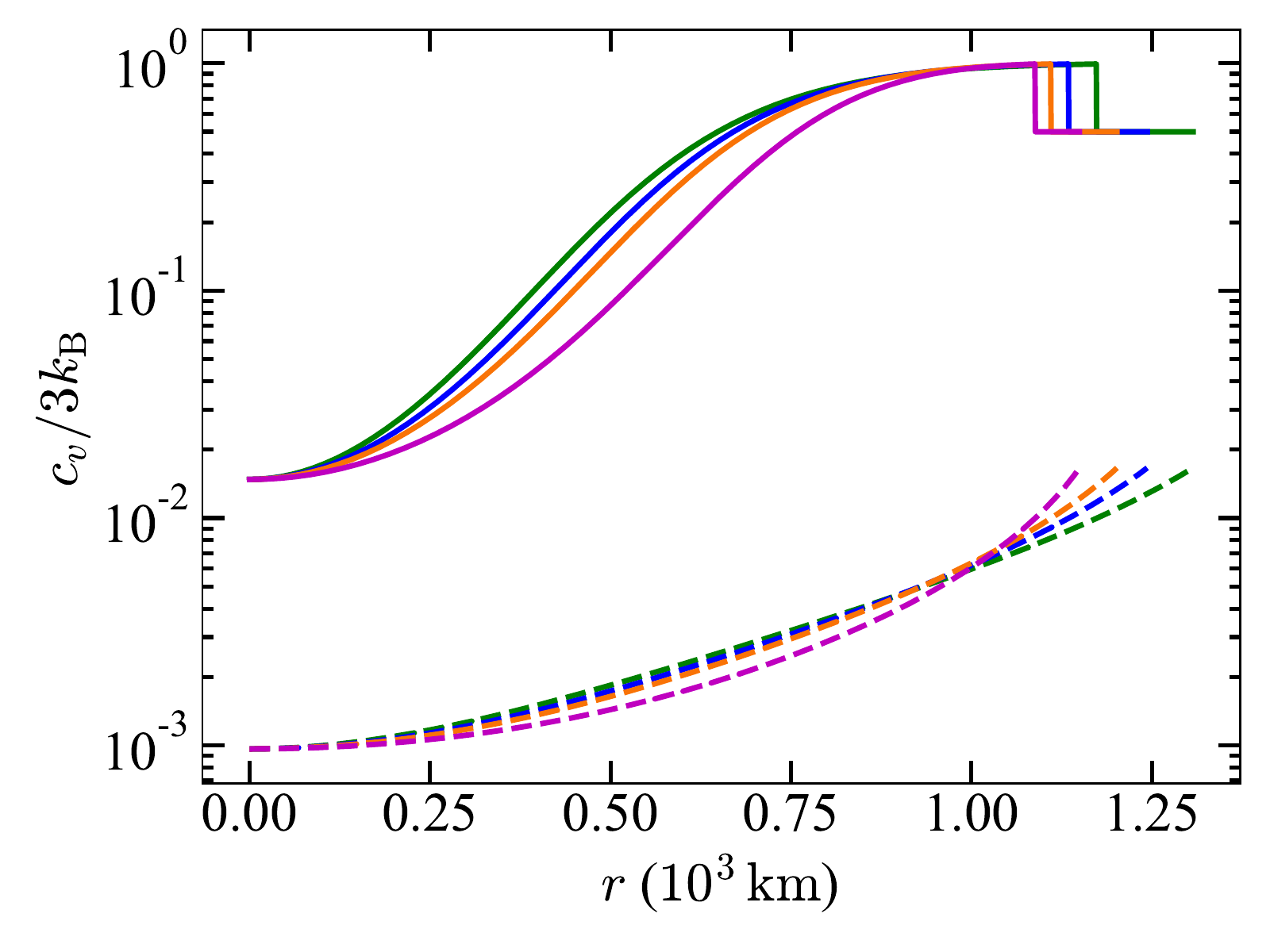}
    }
    \caption{Variation of $c_v$ inside a WD with $\rho_\mathrm{c}=10^{10}\rm\,g\,cm^{-3}$. Dashed lines represent $c_v^\text{el}$ and solid lines represent $c_v^\text{ion}$. Green, blue, orange, and magenta lines represent $\alpha=0$, $-2\times10^{13}$, $-4\times10^{13}$, and $-10^{14}\rm\,cm^2$ respectively.}
    \label{Fig: cv_compare}
\end{figure}

\begin{figure}[htpb]
	\centering
	\includegraphics[scale=0.5]{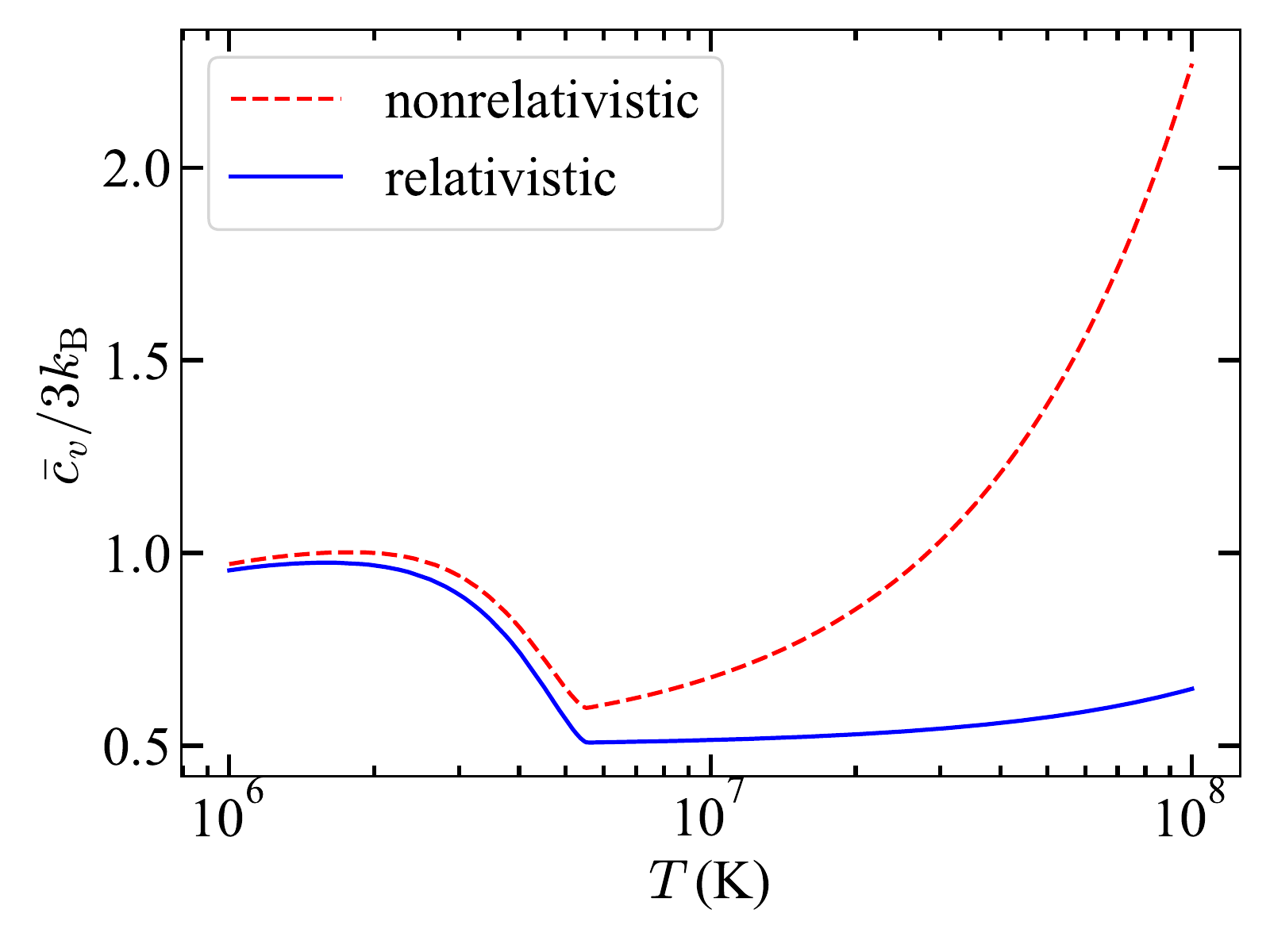}
	\caption{$\bar{c}_v$ as a function of $T$ for carbon WDs for $M=0.36\,M_\odot$ with $\alpha=0$. The red-dashed curve is obtained using the non-relativistic $E_\mathrm{F}-p_\mathrm{F}$ relation while the blue solid curve is for the relativistic one given by Eq.~\eqref{Eq: relativistic EF-pF}. Comparing the red-dashed curve with Fig.~6(a) of the reference~\cite{koester1972outer}, we notice similar behavior.}
	\label{cvn}
\end{figure}

\begin{figure}[htpb]
	\centering
	\includegraphics[scale=0.5]{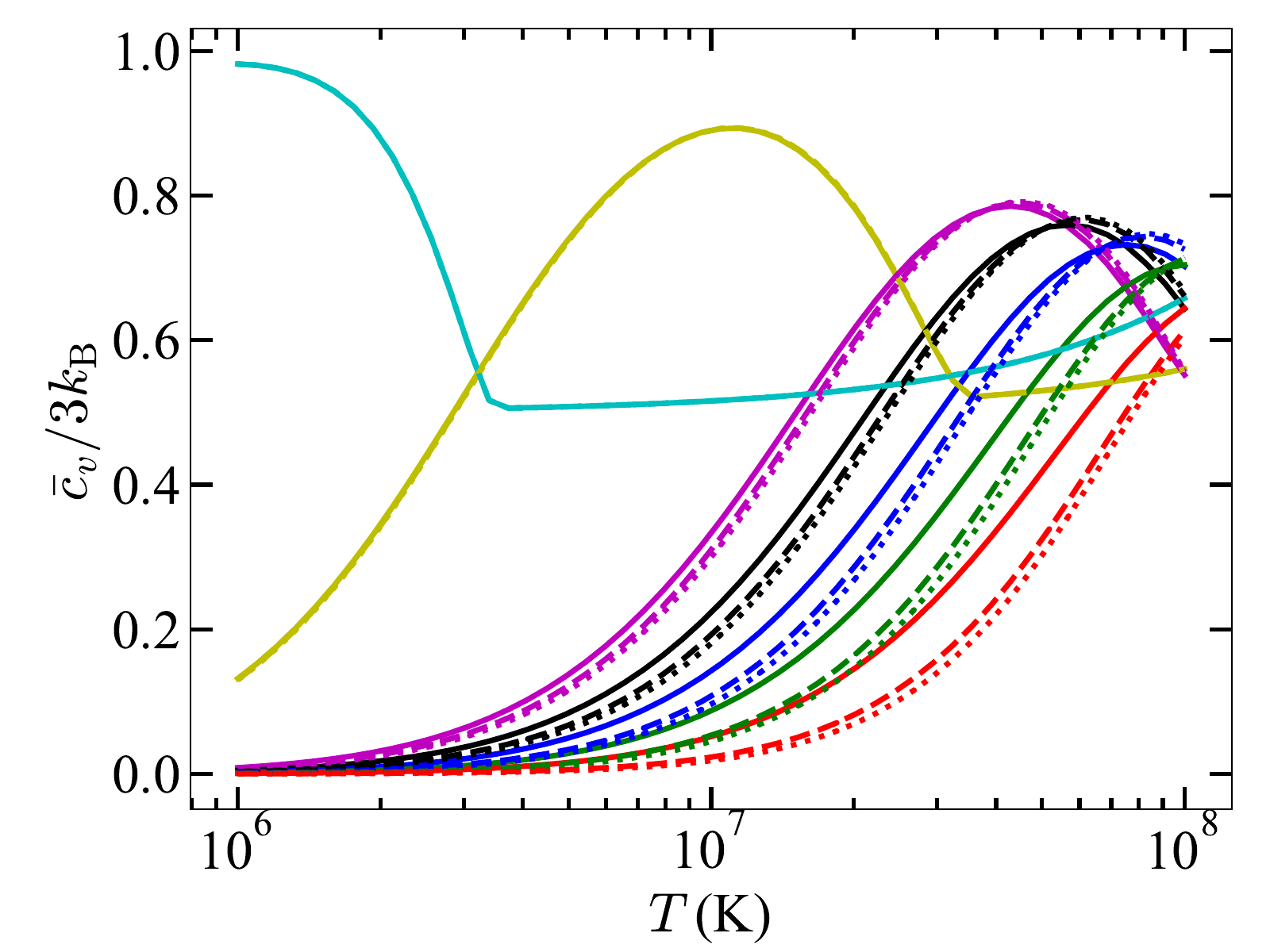}
	\caption{$\bar{c}_v$ as a function of $T$ for different carbon WDs. Solid lines represent conventional WDs under the Newtonian gravity while dashed and dotted lines represent modified gravity inspired WDs with $\alpha=-2\times10^{13}\rm\,cm^2$ and $\alpha=-3\times10^{13}\rm\,cm^2$, respectively. WDs with $\rho_\mathrm{c}=2.0\times10^5\rm\,g\,cm^{-3}$, $2.0\times10^8\rm\,g\,cm^{-3}$, $6.3\times10^9\rm\,g\,cm^{-3}$, $1.3\times10^{10}\rm\,g\,cm^{-3}$, $2.5\times10^{10}\rm\,g\,cm^{-3}$, $5.0\times10^{10}\rm\,g\,cm^{-3}$, and $10^{11}\rm\,g\,cm^{-3}$ are represented by cyan, yellow, magenta, black, blue, green, and red lines, respectively. 
 }
	\label{Fig: CV_T}
\end{figure}

On the other hand, $c_v^\text{ion}$, as already mentioned, depends on the crystallization properties. More specifically, it depends on the critical value of the ratio of Coulomb to thermal energy, denoted by $\Gamma$. Different computations consider that the crystallization may start at different critical $\Gamma$, denoted by $\Gamma_m\in\{60,125\}$~\cite{brush1966monte}. If $\Gamma<\Gamma_m$, $c_v^\text{ion}$ takes the mentioned constant value of $(3/2)k_\mathrm{B}$. However, above $\Gamma_m$, it modifies as
\begin{equation}
    c_v^\text{ion}=9k_\mathrm{B} \left(\frac{T}{\Theta_\mathrm{D}}\right)^3 \int_0^{\Theta_\mathrm{D}/T} \frac{x^4 e^x}{\left(e^x-1\right)^2}\dd{x},
\end{equation}
where the Debye temperature is given by
\begin{equation}\label{debye}
    \Theta_\mathrm{D}=0.174\times10^4 \frac{2Z}{A}\sqrt{\rho}.
\end{equation}
Fig.~\ref{debtemp} shows the variation of $\Theta_\mathrm{D}$, whereas Fig.~\ref{Fig: cv_compare} shows a comparison between $c_v^\text{el}$ and $c_v^\text{ion}$ inside a modified gravity induced WD with $\rho_\mathrm{c}=10^{10}\rm\,g\,cm^{-3}$. Note that we show them only for the negative values of $\alpha$. This is because negative $\alpha$ gives a significant deviation from the standard Newtonian curve of WD. Further, the positive values lead to unstable configurations beyond certain densities (the turnback portions of curves in Fig.~\ref{Fig: MR}). Thus, there is no point of considering the WDs lying in the receding branch of the mass--radius curves. Also, the turn around point is very close to the Newtonian curve and so the effective results for positive $\alpha$ are very similar to the Newtonian results. Hence we do not show anything for positive $\alpha$. Moreover, knowing the exact dependency on density and denoting $c_v = c_v^\text{el}+c_v^\text{ion}$, Eq.~\eqref{mean} can be recast as
\begin{equation}
     \bar{c}_v = \frac{1}{\mathcal M}\int_0^{\mathcal{M}} c_v(T,\rho)\dv{m}{r}\dd{r}.
\end{equation}
Fig.~\ref{cvn} shows a comparison between the effect of relativistic and non-relativistic $p_\mathrm{F}-E_\mathrm{F}$ relation on $\Bar{c}_v$. Further, Fig.~\ref{Fig: CV_T} shows the variation of $\bar{c}_v$ as a function of $T$ for carbon WDs using the relativistic $p_\mathrm{F}-E_\mathrm{F}$ relation given by Eq.~\eqref{Eq: relativistic EF-pF}.

\subsection{Latent heat of the crystallization process}
During the crystallization process, one deals with an additional source of energy which should be taken into account in the energy losses. The latent heat released during the crystallization process is assumed to be $q k_\mathrm{B} T$. Hence, the following additional term contributes to the luminosity~\cite{van1968crystallization}
\begin{equation}
    L_q=qk_\mathrm{B}T \dv{(m_s/A m_\mathrm{p})}{t},
\end{equation}
where $m_s$ is the amount of mass that is already crystallized. Let us now rewrite this equation as follows:
\begin{align}\label{LL}
    L_q &= q k_\mathrm{B}T \frac{\mathcal{M}}{A m_\mathrm{p}} \frac{1}{\mathcal{M}} \dv{m_s}{t} \nonumber\\
    &= q k_\mathrm{B}T \frac{\mathcal{M}}{A m_\mathrm{p}} \frac{1}{\mathcal{M}}\dv{m}{r}\dv{r}{\rho}\dv{\rho_s(T)}{T}\dv{T}{t},
\end{align}
where $\rho_s(T)$ is the density of the crystallized mass at a temperature $T$. It is related to the ratio of Coulomb to thermal energy $\Gamma$ by the relation~\cite{koester1972outer}
\begin{equation}
    \Gamma=2.28\times10^5 \frac{Z^2}{A^{1/3}} \frac{\rho_s^{1/3}}{T}.
\end{equation}
When it reaches the critical value for which the crystallization starts, that is, $\Gamma=\Gamma_m$, one can show that
\begin{equation}
    \dv{\rho_s}{T}= \frac{3\rho_s}{T},
\end{equation}
such that the luminosity in Eq.~\eqref{LL} can be rewritten as
\begin{equation}\label{LL2}
    L_q = 3\rho_s q k_\mathrm{B} \frac{\mathcal{M}}{A m_\mathrm{p}} \frac{1}{\mathcal{M}}\dv{m}{r}\dv{r}{\rho}\dv{T}{t}.
\end{equation}
Note that $\dv*{m}{r}$ and $\dv*{\rho}{r}$ are taken at radius $r_*$ where $\rho(r_*)=\rho_s(T)$ is satisfied.

\subsection{Cooling process}

\begin{figure}[t]
	\centering
	\includegraphics[scale=0.5]{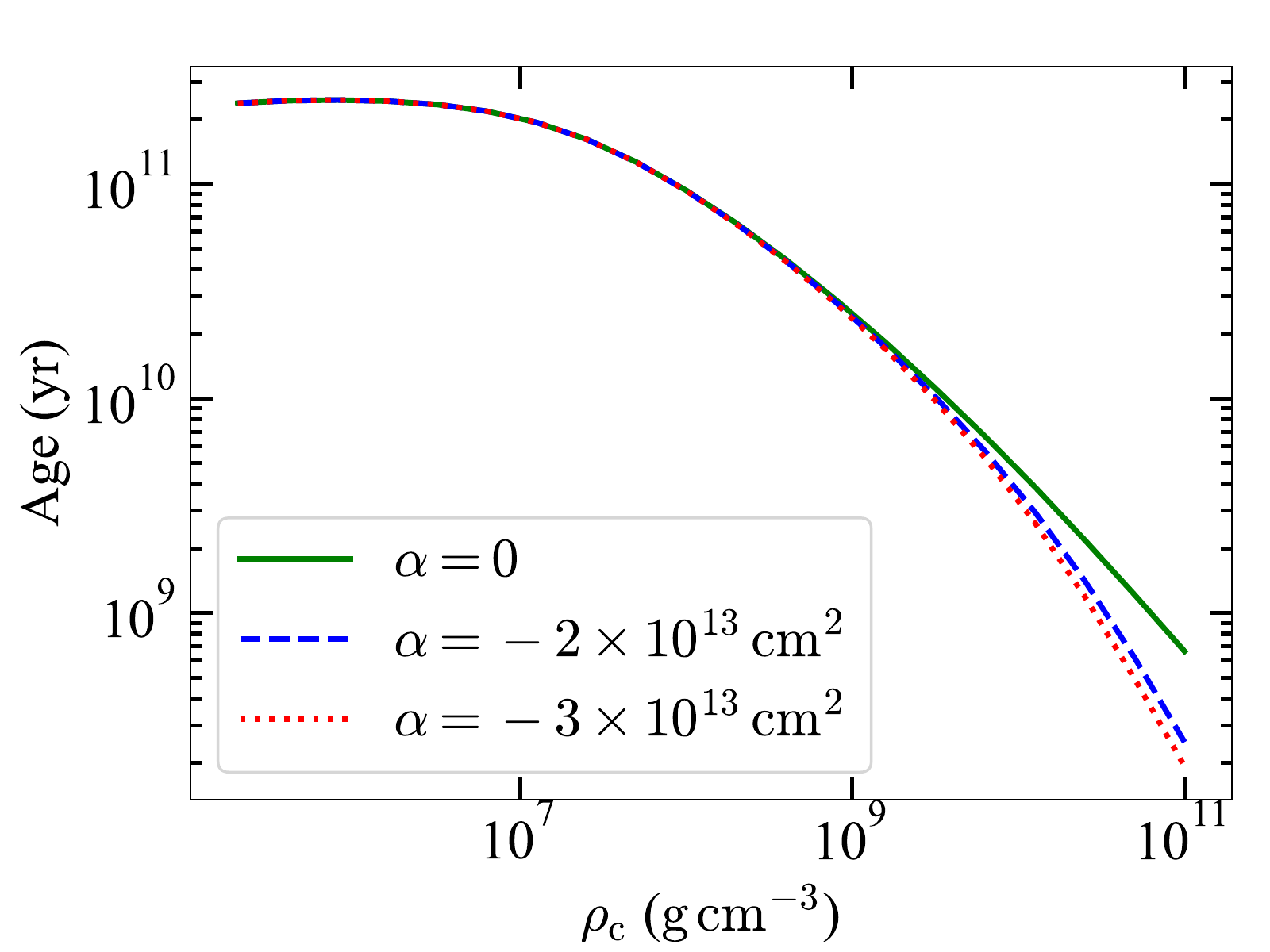}
	\caption{Age of carbon WDs as a function of their central densities, obtained by solving Eq.~\eqref{LU}, when they cool down from $10^8$\,K to $10^6$\,K.}
	\label{Fig: age1}
\end{figure}

\begin{figure}[t]
	\centering
	\includegraphics[scale=0.5]{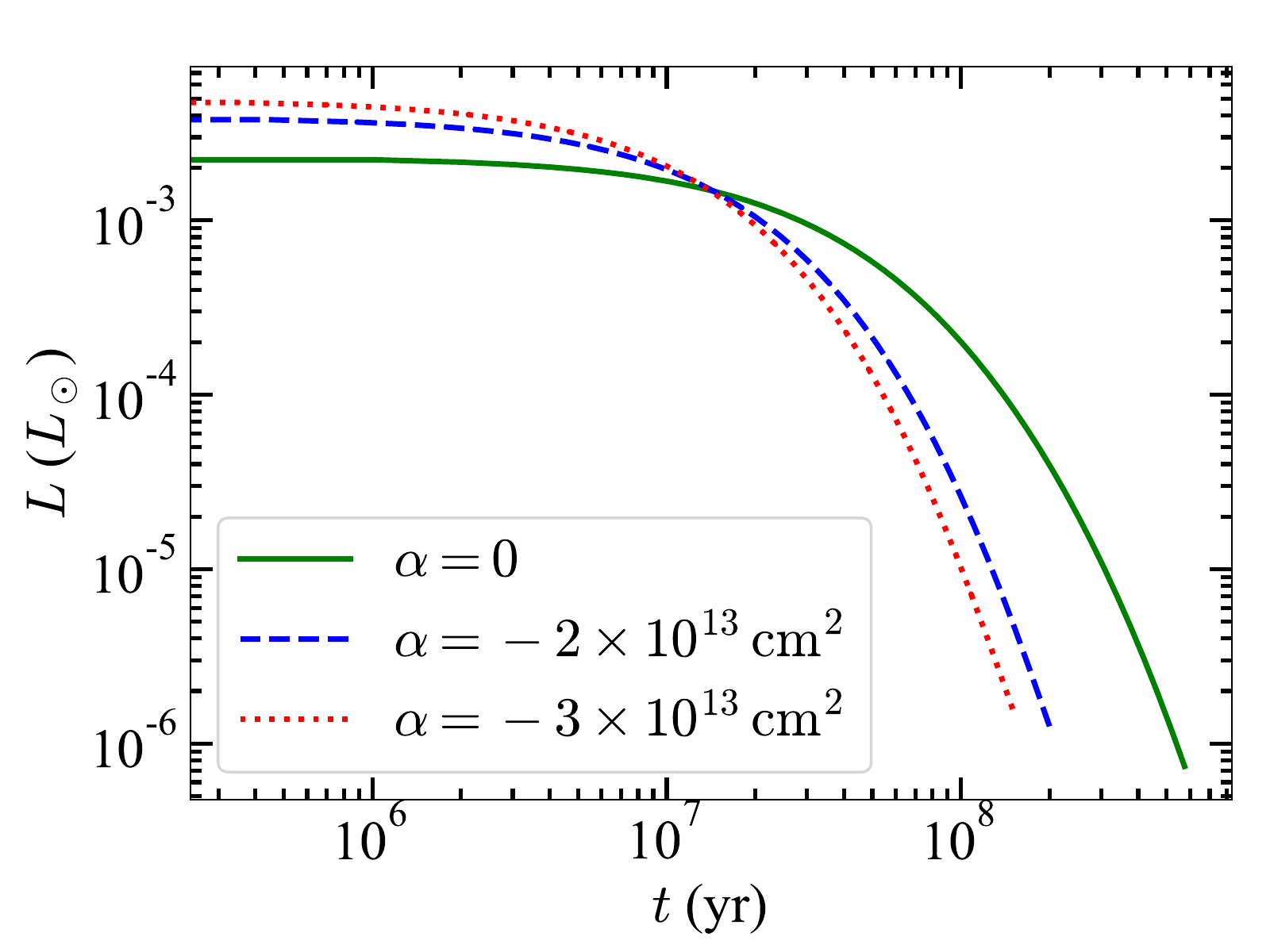}
	\caption{Luminosity as a function of time for carbon WDs with $\rho_\mathrm{c}=10^{11}\rm\, g\,cm^{-3}$. Here the initial luminosity corresponds to a surface temperature of approximately $10^7\rm\,K$. Note that massive WDs fade faster.}
	\label{Fig: Luminosity}
\end{figure}

\begin{figure}[t]
	\centering
	\includegraphics[scale=0.5]{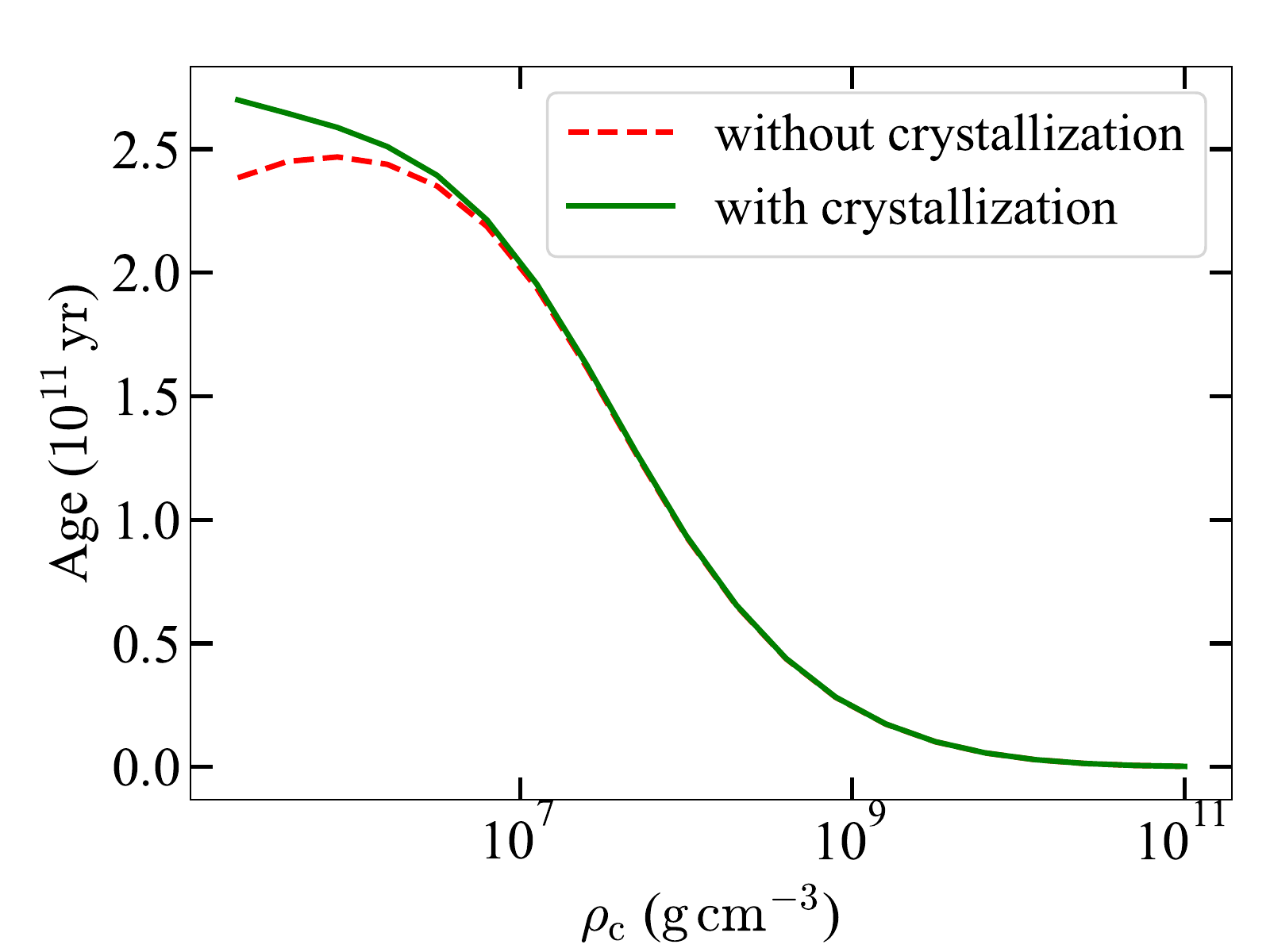}
	\caption{Age of modified gravity induced carbon WDs with $\alpha=-2\times10^{13}\rm\,cm^2$ as a function of their central densities, obtained by solving Eq.~\eqref{LT}, when they cool down from $10^8$\,K to $10^6$\,K.}
	\label{Fig: age2}
\end{figure}

In the presence of crystallization, the final luminosity is the sum of luminosities obtained from Eqs.~\eqref{LU} and~\eqref{LL2}. Thus the modified cooling equation is given by
\begin{equation}\label{LT}
    L = \frac{3k_\mathrm{B}\mathcal{M}}{A m_\mathrm{p}} \left(-\frac{\bar{c}_v}{3k_B} + \rho_s q \frac{1}{\mathcal{M}}\dv{m}{r}\dv{r}{\rho} \right)\dv{T}{t}.
\end{equation}
In our discussion, we consider the initial temperature at $t=0$ to be $10^8$\,K, such that one understands the `age of a white dwarf' as cooling time from this temperature to the present values, which is assumed to be $10^6$\,K. In Fig.~\ref{Fig: age1}, we show the dependency of WD age on the central density for the Newtonian and modified gravity, while Fig.~\ref{Fig: Luminosity} represents how the luminosity fades away with time, also plotted for three values of the parameter $\alpha$. Furthermore, Fig.~\ref{Fig: age2} demonstrates the effects of crystallization in the modified model of gravity. Moreover, the crystallization process we study here is rather expected in more massive WDs because the carbon and oxygen ions in the core solidify when the WD cools down (there exists a low-mass carbon WD because of the cannibalism process of their binary companions, but the lower mass, the slower is the cooling process, so the crystallization would start much later than in more massive objects). Our modeling is still simplified and required further improvements, especially in the case of EoS and taking into account the atmosphere's structure and properties. When we do it, we expect to see the effects of modified gravity much more significant for a wider range of the parameter's value.

\section{Discussion and conclusions}\label{secon}

The aim of the present research was to examine the crystallization process, which is a crucial part of the WDs' cooling phase. We have also taken into account the thermal properties which in our previous work were neglected~\cite{Kalita:2022zki}. Let us note that this problem has not been studied in modified gravity before, and as we have demonstrated, there are a few significant consequences that can have an impact not only on compact stars but also on solid-state physics.

Before discussing our findings, let us briefly summarize our work. Firstly, we have derived the Lane-Emden-Chandrasekhar~(LEC) equation~\eqref{LEC} for the Palatini $f(R)$ gravity, which allows us to obtain the mass, radius, and density profile of a white dwarf whose matter is described by the Chandrasekhar equation of state~\eqref{Chandrasekhar EoS}. This formalism is useful to study the cooling process of white dwarfs since we can express the cooling equations with respect to the solution of the LEC equation.

Further, we have reanalyzed the thermal properties, that is, if, and how, they depend on a model of gravity. Indeed, the mean specific heat in Eq.~\eqref{mean} includes a term related to the phonons' contribution (becoming important when a given material crystallizes), which does depend on the modified gravity model via Debye temperature. This is so because the density profile inside a white dwarf depends on the modified gravity model and the Debye temperature directly depends on the density. Fig.~\ref{debtemp} demonstrates how the Debye temperature in modified gravity differs with respect to the Newtonian one for a given central density $\rho_\mathrm{c}=10^{10}\rm\,g\,cm^{-3}$. We observe that for negative $\alpha$ with super-Chandrasekhar WDs, the Debye temperature is slightly higher in the stellar interior when modified gravity effects are taken into account, while near the surface, it decreases rapidly. This happens because of the fact that the super-Chandrasekhar white dwarfs possess higher mass in comparison to the conventional ones at the same radius (which eventually makes the former to be smaller in size) and because the Debye temperature $\Theta_\mathrm{D} \propto \sqrt{\rho(r)}$, it also follows the same trend of the density profile.

Let us recall that (specific) heat capacity is a property of the material, which depends on temperature and the state of matter. As we have demonstrated here, it also depends on the model of gravity. Fig.~\ref{Fig: cv_compare} shows the specific heat of the electron per ions $c_v^\text{el}$ and of ions $c_v^\text{ion}$. We have seen that at a low temperature and at high density, $c_v^\text{el}$ dominates over $c_v^\text{ion}$, which, however, flips at low density or at high temperature. Moreover, it can be noticed that as we move from the center to the surface, $c_v^\text{ion}$ increases to $3k_\mathrm{B}$ and then sharply decreases to $(3/2)k_\mathrm{B}$ at a particular radius. This is so because, beyond that radius, $\Gamma<\Gamma_m$ is satisfied due to low density. This radius decreases as the temperature increases. In reality, this effect might not be seen with a realistic temperature profile where the surface temperature of a white dwarf is even lower $10^{4}-10^{5}$\,K, and the core temperature is high. Further, we have depicted in Fig.~\ref{cvn} the mean specific heat $\bar{c}_v$ of carbon as a function of temperature for non-relativistic and relativistic Fermi energy in the Newtonian case ($\alpha=0$) in order to check our results with~\cite{koester1972outer}. We see that we have obtained similar behavior as in the mentioned paper. Therefore, in the high temperature, the contribution of the electron per ion dominates in the mean specific heat, which, however, drops rapidly when the temperature decreases till it reaches its crystallization value, and then the mean specific heat rises again up to the maximum value at $\bar v=3k_\text{B}$. It further decreases with temperature since most of the WD's mass cools down below the Debye temperature. On the other hand, we have also shown the relativistic counterparts for the Fermi energy in the same figure, which resulted in slightly different shapes than the Newtonian case. Generally, the decrease in temperature is not so steep as in the non-relativistic case. Notice that relativistic expressions are more relevant for high-density regimes although we also notice a change in the low one.

Fig.~\ref{Fig: CV_T} demonstrates the mean specific heat for a given central density for Newtonian and modified gravity. As we see, the effect of modified gravity is more prominent at high densities, as expected from the mass--radius curve in Fig.~\ref{Fig: MR}. In modified gravity, the mean specific heat reaches lower values than in the Newtonian case for the same temperature. This is exactly the reason why WDs cool down faster in Palatini gravity with respect to the Newtonian one.

Finally, we have reexamined all necessary ingredients to improve our model of the WDs' cooling process. In our previous work~\cite{Kalita:2022zki}, we studied a very simple cooling model with only the thermal energy of the star taken into account. Our current improvement is to include its dependence on the Debye temperature and the ratio of Coulomb to thermal energy as well as incorporating the crystallization process into the cooling model. Thus, we have calculated the latent heat energy resulting from these processes, which contributes to the luminosity. Fig.~\ref{Fig: age1} shows the cooling age of the white dwarfs under modified gravity. Because modified gravity affects the dense WDs, the cooling process is shortened with respect to the Newtonian model. Fig.~\ref{Fig: Luminosity} shows the surface luminosity of the WD as a function of time. The more massive the WD, the quicker decrease in luminosity. Moreover, Fig.~\ref{Fig: age2} shows the cooling timescale when the crystallization is taken into account in the internal processes of the WD. Notice that in general, crystallization prolongs the full cooling process because of the presence of extra energy in the form of latent heat, which needs to be radiated away from the star's surface. This happens to be true irrespective of the gravity model and in this figure, we only show the case with $\alpha=-2\times10^{13}\rm\,cm^2$. However, as already discussed, modified gravity shortens the cooling process. This is a wanted feature in order to explain white dwarfs ``older than the Universe''. Notice that the effect is even more prominent in the case of low-mass WDs, making that in the case of Newtonian physics one can deal with extremely old WDs~\cite{masuda2019self} since none of the well-known and accepted scenarios~\cite{laughlin1997end} was helpful to explain this peculiar phenomenon. Palatini gravity and other theories modifying GR and its non-relativistic limit can provide answers to this issue.
 
We also want to mention a few assumptions we have considered, which should be taken into account in order to improve the cooling models. Firstly, we have studied white dwarfs and their cooling processes in the framework of the non-relativistic limit of gravitational proposals: Newtonian and non-relativistic limit of Palatini $f(R)$ gravity. This is so because generally white dwarfs are big in size and so Newtonian treatment is valid as opposed to the case for small objects like neutron stars. However, it was shown that relativistic effects do also have consequences on the physics of these objects, therefore our next step could be to consider full relativistic theories~\cite{cohen1969treatment,perot2022tidal}. Moreover, because we have examined a spherically symmetric star, we have neglected rotation~\cite{boshkayev2012general} and magnetic fields~\cite{ferrario2020magnetic,mukhopadhyay2021modified}, which can have also a crucial contribution to the observational features and internal properties of the white dwarfs like the pulsation. Apart from it, we have used the Chandrasekhar equation of state which is temperature independent; thus a more realistic matter description will be also required in our next steps~\cite{rotondo2011relativistic,2014PhRvC..89a5801D}. Moreover, the effects of gravity on the equation of state should be also taken into account in order to deal with a fully consistent equation describing a star at the statistical equilibrium~\cite{Wojnar:2022dvo}. 

Let us notice that it is the first time in the literature that the influence of gravity on the specific heat, Debye temperature, and crystallization process has been reported. This property can also be significant in the case of solid-state physics as well as in Earth science. Recently, there were experiments performed in laboratories in which the extreme conditions of the Earth's core were recreated~\cite{merkel2021femtosecond}. It allowed studying the properties and behavior of iron, the main element of the terrestrial planets' cores, under high pressures and temperatures. Although the modified effects can be clearly neglected in weak fields, such as while we deal with the Earth's laboratories~\cite{Kozak:2021ghd}, they can be important in the Earth's and stellar's interiors, as evident in the works of some of us~\cite{Kozak:2021zva,Wojnar:2021xbr,Kozak:2021fjy} (see review, e.g.~\cite{Olmo:2019flu}). It would be intriguing to use our findings to understand the gravitational interaction in dense environments because modified gravity was shown to affect the Earth's core description~\cite{Kozak:2021fjy,Kozak:2021ghd,Kozak:2021zva}. Notice that the microphysics description of the materials, which the Earth composes of, in the given range of temperature and pressure is better understood, providing higher accuracy and comparison with the seismic~\cite{dziewonski1981preliminary} or neutrino~\cite{donini2019neutrino} data. Apart from this, taking some of the mentioned physics into account\footnote{Mainly a more realistic equation of state for interior and envelope, full relativistic description of those objects, and rotation}, our findings can be used to test the existing proposals of gravitational theory against the growing data of white dwarfs~\cite{biesiada2004new,Saltas2018white}. We are working on these ideas and will present the results in future works.

\section*{Acknowledgements}
SK would like to acknowledge support from the South African Research Chairs Initiative of the Department of Science and Technology and the National Research Foundation. AW acknowledges financial support from MICINN (Spain) {\it Ayuda Juan de la Cierva - incorporac\'ion} 2020 No. IJC2020-044751-I and by the EU through the European Regional Development Fund CoE program TK133 ``The Dark Side of the Universe.''

\bibliographystyle{apsrev4-1}
\bibliography{biblio}

\begin{thebibliography}{78}%
\makeatletter
\providecommand \@ifxundefined [1]{%
 \@ifx{#1\undefined}
}%
\providecommand \@ifnum [1]{%
 \ifnum #1\expandafter \@firstoftwo
 \else \expandafter \@secondoftwo
 \fi
}%
\providecommand \@ifx [1]{%
 \ifx #1\expandafter \@firstoftwo
 \else \expandafter \@secondoftwo
 \fi
}%
\providecommand \natexlab [1]{#1}%
\providecommand \enquote  [1]{``#1''}%
\providecommand \bibnamefont  [1]{#1}%
\providecommand \bibfnamefont [1]{#1}%
\providecommand \citenamefont [1]{#1}%
\providecommand \href@noop [0]{\@secondoftwo}%
\providecommand \href [0]{\begingroup \@sanitize@url \@href}%
\providecommand \@href[1]{\@@startlink{#1}\@@href}%
\providecommand \@@href[1]{\endgroup#1\@@endlink}%
\providecommand \@sanitize@url [0]{\catcode `\\12\catcode `\$12\catcode
  `\&12\catcode `\#12\catcode `\^12\catcode `\_12\catcode `\%12\relax}%
\providecommand \@@startlink[1]{}%
\providecommand \@@endlink[0]{}%
\providecommand \url  [0]{\begingroup\@sanitize@url \@url }%
\providecommand \@url [1]{\endgroup\@href {#1}{\urlprefix }}%
\providecommand \urlprefix  [0]{URL }%
\providecommand \Eprint [0]{\href }%
\providecommand \doibase [0]{http://dx.doi.org/}%
\providecommand \selectlanguage [0]{\@gobble}%
\providecommand \bibinfo  [0]{\@secondoftwo}%
\providecommand \bibfield  [0]{\@secondoftwo}%
\providecommand \translation [1]{[#1]}%
\providecommand \BibitemOpen [0]{}%
\providecommand \bibitemStop [0]{}%
\providecommand \bibitemNoStop [0]{.\EOS\space}%
\providecommand \EOS [0]{\spacefactor3000\relax}%
\providecommand \BibitemShut  [1]{\csname bibitem#1\endcsname}%
\let\auto@bib@innerbib\@empty
\bibitem [{\citenamefont {{Shapiro}}\ and\ \citenamefont
  {{Teukolsky}}(1986)}]{1986bhwd.book.....S}%
  \BibitemOpen
  \bibfield  {author} {\bibinfo {author} {\bibfnamefont {S.~L.}\ \bibnamefont
  {{Shapiro}}}\ and\ \bibinfo {author} {\bibfnamefont {S.~A.}\ \bibnamefont
  {{Teukolsky}}},\ }\href@noop {} {\emph {\bibinfo {title} {{Black Holes, White
  Dwarfs and Neutron Stars: The Physics of Compact Objects}}}}\ (\bibinfo
  {publisher} {Wiley-VCH, New York},\ \bibinfo {year} {1986})\BibitemShut
  {NoStop}%
\bibitem [{\citenamefont {{Lauffer}}\ \emph {et~al.}(2018)\citenamefont
  {{Lauffer}}, \citenamefont {{Romero}},\ and\ \citenamefont
  {{Kepler}}}]{2018MNRAS.480.1547L}%
  \BibitemOpen
  \bibfield  {author} {\bibinfo {author} {\bibfnamefont {G.~R.}\ \bibnamefont
  {{Lauffer}}}, \bibinfo {author} {\bibfnamefont {A.~D.}\ \bibnamefont
  {{Romero}}}, \ and\ \bibinfo {author} {\bibfnamefont {S.~O.}\ \bibnamefont
  {{Kepler}}},\ }\href {\doibase 10.1093/mnras/sty1925} {\bibfield  {journal}
  {\bibinfo  {journal} {\mnras}\ }\textbf {\bibinfo {volume} {480}},\ \bibinfo
  {pages} {1547} (\bibinfo {year} {2018})}\BibitemShut {NoStop}%
\bibitem [{\citenamefont {{Caiazzo}}\ \emph {et~al.}(2021)\citenamefont
  {{Caiazzo}} \emph {et~al.}}]{2021Natur.595...39C}%
  \BibitemOpen
  \bibfield  {author} {\bibinfo {author} {\bibfnamefont {I.}~\bibnamefont
  {{Caiazzo}}} \emph {et~al.},\ }\href {\doibase 10.1038/s41586-021-03615-y}
  {\bibfield  {journal} {\bibinfo  {journal} {\nat}\ }\textbf {\bibinfo
  {volume} {595}},\ \bibinfo {pages} {39} (\bibinfo {year} {2021})}\BibitemShut
  {NoStop}%
\bibitem [{\citenamefont {{Anguiano}}\ \emph {et~al.}(2022)\citenamefont
  {{Anguiano}} \emph {et~al.}}]{2022AJ....164..126A}%
  \BibitemOpen
  \bibfield  {author} {\bibinfo {author} {\bibfnamefont {B.}~\bibnamefont
  {{Anguiano}}} \emph {et~al.},\ }\href {\doibase 10.3847/1538-3881/ac8357}
  {\bibfield  {journal} {\bibinfo  {journal} {\aj}\ }\textbf {\bibinfo {volume}
  {164}},\ \bibinfo {eid} {126} (\bibinfo {year} {2022})}\BibitemShut {NoStop}%
\bibitem [{\citenamefont {{Chandrasekhar}}(1935)}]{1935MNRAS..95..207C}%
  \BibitemOpen
  \bibfield  {author} {\bibinfo {author} {\bibfnamefont {S.}~\bibnamefont
  {{Chandrasekhar}}},\ }\href {\doibase 10.1093/mnras/95.3.207} {\bibfield
  {journal} {\bibinfo  {journal} {\mnras}\ }\textbf {\bibinfo {volume} {95}},\
  \bibinfo {pages} {207} (\bibinfo {year} {1935})}\BibitemShut {NoStop}%
\bibitem [{\citenamefont {{Nomoto}}\ \emph {et~al.}(1997)\citenamefont
  {{Nomoto}}, \citenamefont {{Iwamoto}},\ and\ \citenamefont
  {{Kishimoto}}}]{1997Sci...276.1378N}%
  \BibitemOpen
  \bibfield  {author} {\bibinfo {author} {\bibfnamefont {K.}~\bibnamefont
  {{Nomoto}}}, \bibinfo {author} {\bibfnamefont {K.}~\bibnamefont {{Iwamoto}}},
  \ and\ \bibinfo {author} {\bibfnamefont {N.}~\bibnamefont {{Kishimoto}}},\
  }\href {\doibase 10.1126/science.276.5317.1378} {\bibfield  {journal}
  {\bibinfo  {journal} {Science}\ }\textbf {\bibinfo {volume} {276}},\ \bibinfo
  {pages} {1378} (\bibinfo {year} {1997})}\BibitemShut {NoStop}%
\bibitem [{\citenamefont {{Lieb}}\ and\ \citenamefont
  {{Yau}}(1987)}]{1987ApJ...323..140L}%
  \BibitemOpen
  \bibfield  {author} {\bibinfo {author} {\bibfnamefont {E.~H.}\ \bibnamefont
  {{Lieb}}}\ and\ \bibinfo {author} {\bibfnamefont {H.-T.}\ \bibnamefont
  {{Yau}}},\ }\href {\doibase 10.1086/165813} {\bibfield  {journal} {\bibinfo
  {journal} {\apj}\ }\textbf {\bibinfo {volume} {323}},\ \bibinfo {pages} {140}
  (\bibinfo {year} {1987})}\BibitemShut {NoStop}%
\bibitem [{\citenamefont {{Howell}}\ \emph {et~al.}(2006)\citenamefont
  {{Howell}} \emph {et~al.}}]{2006Natur.443..308H}%
  \BibitemOpen
  \bibfield  {author} {\bibinfo {author} {\bibfnamefont {D.~A.}\ \bibnamefont
  {{Howell}}} \emph {et~al.},\ }\href {\doibase 10.1038/nature05103} {\bibfield
   {journal} {\bibinfo  {journal} {\nat}\ }\textbf {\bibinfo {volume} {443}},\
  \bibinfo {pages} {308} (\bibinfo {year} {2006})}\BibitemShut {NoStop}%
\bibitem [{\citenamefont {{Scalzo}}\ \emph {et~al.}(2010)\citenamefont
  {{Scalzo}} \emph {et~al.}}]{2010ApJ...713.1073S}%
  \BibitemOpen
  \bibfield  {author} {\bibinfo {author} {\bibfnamefont {R.~A.}\ \bibnamefont
  {{Scalzo}}} \emph {et~al.},\ }\href {\doibase 10.1088/0004-637X/713/2/1073}
  {\bibfield  {journal} {\bibinfo  {journal} {\apj}\ }\textbf {\bibinfo
  {volume} {713}},\ \bibinfo {pages} {1073} (\bibinfo {year}
  {2010})}\BibitemShut {NoStop}%
\bibitem [{\citenamefont {{Yamanaka}}\ \emph {et~al.}(2009)\citenamefont
  {{Yamanaka}} \emph {et~al.}}]{2009ApJ...707L.118Y}%
  \BibitemOpen
  \bibfield  {author} {\bibinfo {author} {\bibfnamefont {M.}~\bibnamefont
  {{Yamanaka}}} \emph {et~al.},\ }\href {\doibase 10.1088/0004-637X/707/2/L118}
  {\bibfield  {journal} {\bibinfo  {journal} {\apjl}\ }\textbf {\bibinfo
  {volume} {707}},\ \bibinfo {pages} {L118} (\bibinfo {year}
  {2009})}\BibitemShut {NoStop}%
\bibitem [{\citenamefont {{Silverman}}\ \emph {et~al.}(2013)\citenamefont
  {{Silverman}}, \citenamefont {{Ganeshalingam}},\ and\ \citenamefont
  {{Filippenko}}}]{2013MNRAS.430.1030S}%
  \BibitemOpen
  \bibfield  {author} {\bibinfo {author} {\bibfnamefont {J.~M.}\ \bibnamefont
  {{Silverman}}}, \bibinfo {author} {\bibfnamefont {M.}~\bibnamefont
  {{Ganeshalingam}}}, \ and\ \bibinfo {author} {\bibfnamefont {A.~V.}\
  \bibnamefont {{Filippenko}}},\ }\href {\doibase 10.1093/mnras/sts674}
  {\bibfield  {journal} {\bibinfo  {journal} {\mnras}\ }\textbf {\bibinfo
  {volume} {430}},\ \bibinfo {pages} {1030} (\bibinfo {year}
  {2013})}\BibitemShut {NoStop}%
\bibitem [{\citenamefont {{Filippenko}}\ \emph {et~al.}(1992)\citenamefont
  {{Filippenko}} \emph {et~al.}}]{1992AJ....104.1543F}%
  \BibitemOpen
  \bibfield  {author} {\bibinfo {author} {\bibfnamefont {A.~V.}\ \bibnamefont
  {{Filippenko}}} \emph {et~al.},\ }\href {\doibase 10.1086/116339} {\bibfield
  {journal} {\bibinfo  {journal} {\aj}\ }\textbf {\bibinfo {volume} {104}},\
  \bibinfo {pages} {1543} (\bibinfo {year} {1992})}\BibitemShut {NoStop}%
\bibitem [{\citenamefont {{Turatto}}\ \emph {et~al.}(1998)\citenamefont
  {{Turatto}} \emph {et~al.}}]{1998AJ....116.2431T}%
  \BibitemOpen
  \bibfield  {author} {\bibinfo {author} {\bibfnamefont {M.}~\bibnamefont
  {{Turatto}}} \emph {et~al.},\ }\href {\doibase 10.1086/300622} {\bibfield
  {journal} {\bibinfo  {journal} {\aj}\ }\textbf {\bibinfo {volume} {116}},\
  \bibinfo {pages} {2431} (\bibinfo {year} {1998})}\BibitemShut {NoStop}%
\bibitem [{\citenamefont {{Taubenberger}}\ \emph {et~al.}(2008)\citenamefont
  {{Taubenberger}} \emph {et~al.}}]{2008MNRAS.385...75T}%
  \BibitemOpen
  \bibfield  {author} {\bibinfo {author} {\bibfnamefont {S.}~\bibnamefont
  {{Taubenberger}}} \emph {et~al.},\ }\href {\doibase
  10.1111/j.1365-2966.2008.12843.x} {\bibfield  {journal} {\bibinfo  {journal}
  {\mnras}\ }\textbf {\bibinfo {volume} {385}},\ \bibinfo {pages} {75}
  (\bibinfo {year} {2008})}\BibitemShut {NoStop}%
\bibitem [{\citenamefont {{Khokhlov}}\ \emph {et~al.}(1993)\citenamefont
  {{Khokhlov}}, \citenamefont {{Mueller}},\ and\ \citenamefont
  {{Hoeflich}}}]{1993A&A...270..223K}%
  \BibitemOpen
  \bibfield  {author} {\bibinfo {author} {\bibfnamefont {A.}~\bibnamefont
  {{Khokhlov}}}, \bibinfo {author} {\bibfnamefont {E.}~\bibnamefont
  {{Mueller}}}, \ and\ \bibinfo {author} {\bibfnamefont {P.}~\bibnamefont
  {{Hoeflich}}},\ }\href@noop {} {\bibfield  {journal} {\bibinfo  {journal}
  {\aap}\ }\textbf {\bibinfo {volume} {270}},\ \bibinfo {pages} {223} (\bibinfo
  {year} {1993})}\BibitemShut {NoStop}%
\bibitem [{\citenamefont {{Buchdahl}}(1970)}]{1970MNRAS.150....1B}%
  \BibitemOpen
  \bibfield  {author} {\bibinfo {author} {\bibfnamefont {H.~A.}\ \bibnamefont
  {{Buchdahl}}},\ }\href {\doibase 10.1093/mnras/150.1.1} {\bibfield  {journal}
  {\bibinfo  {journal} {\mnras}\ }\textbf {\bibinfo {volume} {150}},\ \bibinfo
  {pages} {1} (\bibinfo {year} {1970})}\BibitemShut {NoStop}%
\bibitem [{\citenamefont {{Capozziello}}\ and\ \citenamefont {{de
  Laurentis}}(2011)}]{2011PhR...509..167C}%
  \BibitemOpen
  \bibfield  {author} {\bibinfo {author} {\bibfnamefont {S.}~\bibnamefont
  {{Capozziello}}}\ and\ \bibinfo {author} {\bibfnamefont {M.}~\bibnamefont
  {{de Laurentis}}},\ }\href {\doibase 10.1016/j.physrep.2011.09.003}
  {\bibfield  {journal} {\bibinfo  {journal} {\physrep}\ }\textbf {\bibinfo
  {volume} {509}},\ \bibinfo {pages} {167} (\bibinfo {year}
  {2011})}\BibitemShut {NoStop}%
\bibitem [{\citenamefont {Sotiriou}\ and\ \citenamefont
  {Faraoni}(2010)}]{sotiriou2010f}%
  \BibitemOpen
  \bibfield  {author} {\bibinfo {author} {\bibfnamefont {T.~P.}\ \bibnamefont
  {Sotiriou}}\ and\ \bibinfo {author} {\bibfnamefont {V.}~\bibnamefont
  {Faraoni}},\ }\href {\doibase 10.1103/RevModPhys.82.451} {\bibfield
  {journal} {\bibinfo  {journal} {Reviews of Modern Physics}\ }\textbf
  {\bibinfo {volume} {82}},\ \bibinfo {pages} {451} (\bibinfo {year}
  {2010})}\BibitemShut {NoStop}%
\bibitem [{\citenamefont {{Fay}}\ \emph {et~al.}(2007)\citenamefont {{Fay}},
  \citenamefont {{Tavakol}},\ and\ \citenamefont
  {{Tsujikawa}}}]{2007PhRvD..75f3509F}%
  \BibitemOpen
  \bibfield  {author} {\bibinfo {author} {\bibfnamefont {S.}~\bibnamefont
  {{Fay}}}, \bibinfo {author} {\bibfnamefont {R.}~\bibnamefont {{Tavakol}}}, \
  and\ \bibinfo {author} {\bibfnamefont {S.}~\bibnamefont {{Tsujikawa}}},\
  }\href {\doibase 10.1103/PhysRevD.75.063509} {\bibfield  {journal} {\bibinfo
  {journal} {\prd}\ }\textbf {\bibinfo {volume} {75}},\ \bibinfo {eid} {063509}
  (\bibinfo {year} {2007})}\BibitemShut {NoStop}%
\bibitem [{\citenamefont
  {{Sotiriou}}(2006{\natexlab{a}})}]{2006CQGra..23.1253S}%
  \BibitemOpen
  \bibfield  {author} {\bibinfo {author} {\bibfnamefont {T.~P.}\ \bibnamefont
  {{Sotiriou}}},\ }\href {\doibase 10.1088/0264-9381/23/4/012} {\bibfield
  {journal} {\bibinfo  {journal} {Classical and Quantum Gravity}\ }\textbf
  {\bibinfo {volume} {23}},\ \bibinfo {pages} {1253} (\bibinfo {year}
  {2006}{\natexlab{a}})}\BibitemShut {NoStop}%
\bibitem [{\citenamefont
  {{Sotiriou}}(2006{\natexlab{b}})}]{2006PhRvD..73f3515S}%
  \BibitemOpen
  \bibfield  {author} {\bibinfo {author} {\bibfnamefont {T.~P.}\ \bibnamefont
  {{Sotiriou}}},\ }\href {\doibase 10.1103/PhysRevD.73.063515} {\bibfield
  {journal} {\bibinfo  {journal} {\prd}\ }\textbf {\bibinfo {volume} {73}},\
  \bibinfo {eid} {063515} (\bibinfo {year} {2006}{\natexlab{b}})}\BibitemShut
  {NoStop}%
\bibitem [{\citenamefont {{Nojiri}}\ and\ \citenamefont
  {{Odintsov}}(2004)}]{2004GReGr..36.1765N}%
  \BibitemOpen
  \bibfield  {author} {\bibinfo {author} {\bibfnamefont {S.}~\bibnamefont
  {{Nojiri}}}\ and\ \bibinfo {author} {\bibfnamefont {S.~D.}\ \bibnamefont
  {{Odintsov}}},\ }\href {\doibase 10.1023/B:GERG.0000035950.40718.48}
  {\bibfield  {journal} {\bibinfo  {journal} {General Relativity and
  Gravitation}\ }\textbf {\bibinfo {volume} {36}},\ \bibinfo {pages} {1765}
  (\bibinfo {year} {2004})}\BibitemShut {NoStop}%
\bibitem [{\citenamefont {{Amarzguioui}}\ \emph {et~al.}(2006)\citenamefont
  {{Amarzguioui}} \emph {et~al.}}]{2006A&A...454..707A}%
  \BibitemOpen
  \bibfield  {author} {\bibinfo {author} {\bibfnamefont {M.}~\bibnamefont
  {{Amarzguioui}}} \emph {et~al.},\ }\href {\doibase
  10.1051/0004-6361:20064994} {\bibfield  {journal} {\bibinfo  {journal}
  {\aap}\ }\textbf {\bibinfo {volume} {454}},\ \bibinfo {pages} {707} (\bibinfo
  {year} {2006})}\BibitemShut {NoStop}%
\bibitem [{\citenamefont {{Borowiec}}\ \emph {et~al.}(2016)\citenamefont
  {{Borowiec}} \emph {et~al.}}]{2016JCAP...01..040B}%
  \BibitemOpen
  \bibfield  {author} {\bibinfo {author} {\bibfnamefont {A.}~\bibnamefont
  {{Borowiec}}} \emph {et~al.},\ }\href {\doibase
  10.1088/1475-7516/2016/01/040} {\bibfield  {journal} {\bibinfo  {journal}
  {\jcap}\ }\textbf {\bibinfo {volume} {2016}},\ \bibinfo {pages} {040}
  (\bibinfo {year} {2016})}\BibitemShut {NoStop}%
\bibitem [{\citenamefont {{Teppa Pannia}}\ \emph {et~al.}(2017)\citenamefont
  {{Teppa Pannia}} \emph {et~al.}}]{2017GReGr..49...25T}%
  \BibitemOpen
  \bibfield  {author} {\bibinfo {author} {\bibfnamefont {F.~A.}\ \bibnamefont
  {{Teppa Pannia}}} \emph {et~al.},\ }\href {\doibase
  10.1007/s10714-016-2182-7} {\bibfield  {journal} {\bibinfo  {journal}
  {General Relativity and Gravitation}\ }\textbf {\bibinfo {volume} {49}},\
  \bibinfo {eid} {25} (\bibinfo {year} {2017})}\BibitemShut {NoStop}%
\bibitem [{\citenamefont {{Herzog}}\ and\ \citenamefont
  {{Sanchis-Alepuz}}(2021)}]{2021EPJC...81..888H}%
  \BibitemOpen
  \bibfield  {author} {\bibinfo {author} {\bibfnamefont {G.}~\bibnamefont
  {{Herzog}}}\ and\ \bibinfo {author} {\bibfnamefont {H.}~\bibnamefont
  {{Sanchis-Alepuz}}},\ }\href {\doibase 10.1140/epjc/s10052-021-09662-z}
  {\bibfield  {journal} {\bibinfo  {journal} {Eur. Phys. J. C}\ }\textbf
  {\bibinfo {volume} {81}},\ \bibinfo {eid} {888} (\bibinfo {year}
  {2021})}\BibitemShut {NoStop}%
\bibitem [{\citenamefont {{Olmo}}\ \emph {et~al.}(2019)\citenamefont {{Olmo}},
  \citenamefont {{Rubiera-Garcia}},\ and\ \citenamefont
  {{Wojnar}}}]{2019PhRvD.100d4020O}%
  \BibitemOpen
  \bibfield  {author} {\bibinfo {author} {\bibfnamefont {G.~J.}\ \bibnamefont
  {{Olmo}}}, \bibinfo {author} {\bibfnamefont {D.}~\bibnamefont
  {{Rubiera-Garcia}}}, \ and\ \bibinfo {author} {\bibfnamefont
  {A.}~\bibnamefont {{Wojnar}}},\ }\href {\doibase 10.1103/PhysRevD.100.044020}
  {\bibfield  {journal} {\bibinfo  {journal} {\prd}\ }\textbf {\bibinfo
  {volume} {100}},\ \bibinfo {eid} {044020} (\bibinfo {year}
  {2019})}\BibitemShut {NoStop}%
\bibitem [{\citenamefont {{Das}}\ and\ \citenamefont
  {{Mukhopadhyay}}(2015)}]{2015JCAP...05..045D}%
  \BibitemOpen
  \bibfield  {author} {\bibinfo {author} {\bibfnamefont {U.}~\bibnamefont
  {{Das}}}\ and\ \bibinfo {author} {\bibfnamefont {B.}~\bibnamefont
  {{Mukhopadhyay}}},\ }\href {\doibase 10.1088/1475-7516/2015/05/045}
  {\bibfield  {journal} {\bibinfo  {journal} {\jcap}\ }\textbf {\bibinfo
  {volume} {5}},\ \bibinfo {eid} {045} (\bibinfo {year} {2015})}\BibitemShut
  {NoStop}%
\bibitem [{\citenamefont {{Kalita}}\ and\ \citenamefont
  {{Mukhopadhyay}}(2018)}]{2018JCAP...09..007K}%
  \BibitemOpen
  \bibfield  {author} {\bibinfo {author} {\bibfnamefont {S.}~\bibnamefont
  {{Kalita}}}\ and\ \bibinfo {author} {\bibfnamefont {B.}~\bibnamefont
  {{Mukhopadhyay}}},\ }\href {\doibase 10.1088/1475-7516/2018/09/007}
  {\bibfield  {journal} {\bibinfo  {journal} {\jcap}\ }\textbf {\bibinfo
  {volume} {9}},\ \bibinfo {eid} {007} (\bibinfo {year} {2018})}\BibitemShut
  {NoStop}%
\bibitem [{\citenamefont {{Kalita}}\ and\ \citenamefont
  {{Sarmah}}(2022)}]{2022PhLB..82736942K}%
  \BibitemOpen
  \bibfield  {author} {\bibinfo {author} {\bibfnamefont {S.}~\bibnamefont
  {{Kalita}}}\ and\ \bibinfo {author} {\bibfnamefont {L.}~\bibnamefont
  {{Sarmah}}},\ }\href {\doibase 10.1016/j.physletb.2022.136942} {\bibfield
  {journal} {\bibinfo  {journal} {Phys. Lett. B}\ }\textbf {\bibinfo {volume}
  {827}},\ \bibinfo {eid} {136942} (\bibinfo {year} {2022})}\BibitemShut
  {NoStop}%
\bibitem [{\citenamefont {{Astashenok}}\ \emph {et~al.}(2022)\citenamefont
  {{Astashenok}}, \citenamefont {{Odintsov}},\ and\ \citenamefont
  {{Oikonomou}}}]{astashenok2022maximal}%
  \BibitemOpen
  \bibfield  {author} {\bibinfo {author} {\bibfnamefont {A.~V.}\ \bibnamefont
  {{Astashenok}}}, \bibinfo {author} {\bibfnamefont {S.~D.}\ \bibnamefont
  {{Odintsov}}}, \ and\ \bibinfo {author} {\bibfnamefont {V.~K.}\ \bibnamefont
  {{Oikonomou}}},\ }\href {\doibase 10.1103/PhysRevD.106.124010} {\bibfield
  {journal} {\bibinfo  {journal} {\prd}\ }\textbf {\bibinfo {volume} {106}},\
  \bibinfo {eid} {124010} (\bibinfo {year} {2022})}\BibitemShut {NoStop}%
\bibitem [{\citenamefont {{Sarmah}}\ \emph {et~al.}(2022)\citenamefont
  {{Sarmah}}, \citenamefont {{Kalita}},\ and\ \citenamefont
  {{Wojnar}}}]{2022PhRvD.105b4028S}%
  \BibitemOpen
  \bibfield  {author} {\bibinfo {author} {\bibfnamefont {L.}~\bibnamefont
  {{Sarmah}}}, \bibinfo {author} {\bibfnamefont {S.}~\bibnamefont {{Kalita}}},
  \ and\ \bibinfo {author} {\bibfnamefont {A.}~\bibnamefont {{Wojnar}}},\
  }\href {\doibase 10.1103/PhysRevD.105.024028} {\bibfield  {journal} {\bibinfo
   {journal} {\prd}\ }\textbf {\bibinfo {volume} {105}},\ \bibinfo {eid}
  {024028} (\bibinfo {year} {2022})}\BibitemShut {NoStop}%
\bibitem [{\citenamefont {{Kalita}}\ and\ \citenamefont
  {{Mukhopadhyay}}(2021)}]{2021ApJ...909...65K}%
  \BibitemOpen
  \bibfield  {author} {\bibinfo {author} {\bibfnamefont {S.}~\bibnamefont
  {{Kalita}}}\ and\ \bibinfo {author} {\bibfnamefont {B.}~\bibnamefont
  {{Mukhopadhyay}}},\ }\href {\doibase 10.3847/1538-4357/abddb8} {\bibfield
  {journal} {\bibinfo  {journal} {\apj}\ }\textbf {\bibinfo {volume} {909}},\
  \bibinfo {eid} {65} (\bibinfo {year} {2021})}\BibitemShut {NoStop}%
\bibitem [{\citenamefont {Olmo}\ \emph {et~al.}(2020)\citenamefont {Olmo},
  \citenamefont {Rubiera-Garcia},\ and\ \citenamefont {Wojnar}}]{Olmo:2019flu}%
  \BibitemOpen
  \bibfield  {author} {\bibinfo {author} {\bibfnamefont {G.~J.}\ \bibnamefont
  {Olmo}}, \bibinfo {author} {\bibfnamefont {D.}~\bibnamefont
  {Rubiera-Garcia}}, \ and\ \bibinfo {author} {\bibfnamefont {A.}~\bibnamefont
  {Wojnar}},\ }\href {\doibase 10.1016/j.physrep.2020.07.001} {\bibfield
  {journal} {\bibinfo  {journal} {Phys. Rept.}\ }\textbf {\bibinfo {volume}
  {876}},\ \bibinfo {pages} {1} (\bibinfo {year} {2020})}\BibitemShut {NoStop}%
\bibitem [{\citenamefont {Saltas}\ and\ \citenamefont
  {Lopes}(2019)}]{saltas2019obtaining}%
  \BibitemOpen
  \bibfield  {author} {\bibinfo {author} {\bibfnamefont {I.~D.}\ \bibnamefont
  {Saltas}}\ and\ \bibinfo {author} {\bibfnamefont {I.}~\bibnamefont {Lopes}},\
  }\href {\doibase 10.1103/PhysRevLett.123.091103} {\bibfield  {journal}
  {\bibinfo  {journal} {\prl}\ }\textbf {\bibinfo {volume} {123}},\ \bibinfo
  {pages} {091103} (\bibinfo {year} {2019})}\BibitemShut {NoStop}%
\bibitem [{\citenamefont {Kozak}\ and\ \citenamefont
  {Wojnar}(2021{\natexlab{a}})}]{Kozak:2021ghd}%
  \BibitemOpen
  \bibfield  {author} {\bibinfo {author} {\bibfnamefont {A.}~\bibnamefont
  {Kozak}}\ and\ \bibinfo {author} {\bibfnamefont {A.}~\bibnamefont {Wojnar}},\
  }\href {\doibase 10.1103/PhysRevD.104.084097} {\bibfield  {journal} {\bibinfo
   {journal} {\prd}\ }\textbf {\bibinfo {volume} {104}},\ \bibinfo {pages}
  {084097} (\bibinfo {year} {2021}{\natexlab{a}})}\BibitemShut {NoStop}%
\bibitem [{\citenamefont {Kozak}\ and\ \citenamefont
  {Wojnar}(2022)}]{Kozak:2021zva}%
  \BibitemOpen
  \bibfield  {author} {\bibinfo {author} {\bibfnamefont {A.}~\bibnamefont
  {Kozak}}\ and\ \bibinfo {author} {\bibfnamefont {A.}~\bibnamefont {Wojnar}},\
  }\href {\doibase 10.1142/S0219887822501572} {\bibfield  {journal} {\bibinfo
  {journal} {Int. J. Geom. Meth. Mod. Phys.}\ }\textbf {\bibinfo {volume}
  {19}},\ \bibinfo {pages} {2250157} (\bibinfo {year} {2022})}\BibitemShut
  {NoStop}%
\bibitem [{\citenamefont {Kozak}\ and\ \citenamefont
  {Wojnar}(2021{\natexlab{b}})}]{Kozak:2021fjy}%
  \BibitemOpen
  \bibfield  {author} {\bibinfo {author} {\bibfnamefont {A.}~\bibnamefont
  {Kozak}}\ and\ \bibinfo {author} {\bibfnamefont {A.}~\bibnamefont {Wojnar}},\
  }\href {\doibase 10.3390/universe8010003} {\bibfield  {journal} {\bibinfo
  {journal} {Universe}\ }\textbf {\bibinfo {volume} {8}},\ \bibinfo {pages} {3}
  (\bibinfo {year} {2021}{\natexlab{b}})}\BibitemShut {NoStop}%
\bibitem [{\citenamefont {Wojnar}(2021{\natexlab{a}})}]{Wojnar:2020frr}%
  \BibitemOpen
  \bibfield  {author} {\bibinfo {author} {\bibfnamefont {A.}~\bibnamefont
  {Wojnar}},\ }\href {\doibase 10.1103/PhysRevD.103.044037} {\bibfield
  {journal} {\bibinfo  {journal} {Phys. Rev. D}\ }\textbf {\bibinfo {volume}
  {103}},\ \bibinfo {pages} {044037} (\bibinfo {year}
  {2021}{\natexlab{a}})}\BibitemShut {NoStop}%
\bibitem [{\citenamefont {Wojnar}(2022)}]{Wojnar:2022txk}%
  \BibitemOpen
  \bibfield  {author} {\bibinfo {author} {\bibfnamefont {A.}~\bibnamefont
  {Wojnar}},\ }\enquote {\bibinfo {title} {{Stellar and substellar objects in
  modified gravity}},}\ \ (\bibinfo {year} {2022})\ \Eprint
  {http://arxiv.org/abs/2205.08160} {arXiv:2205.08160 [gr-qc]} \BibitemShut
  {NoStop}%
\bibitem [{\citenamefont {Kulikov}\ and\ \citenamefont
  {Pronin}(1995)}]{kulikov1995low}%
  \BibitemOpen
  \bibfield  {author} {\bibinfo {author} {\bibfnamefont {I.~K.}\ \bibnamefont
  {Kulikov}}\ and\ \bibinfo {author} {\bibfnamefont {P.~I.}\ \bibnamefont
  {Pronin}},\ }\href {\doibase 10.1007/BF00674065} {\bibfield  {journal}
  {\bibinfo  {journal} {Int. J. Theor. Phys.}\ }\textbf {\bibinfo {volume}
  {34}},\ \bibinfo {pages} {1843} (\bibinfo {year} {1995})}\BibitemShut
  {NoStop}%
\bibitem [{\citenamefont {{Li}}\ \emph {et~al.}(2022)\citenamefont {{Li}},
  \citenamefont {{Guo}}, \citenamefont {{Zhao}},\ and\ \citenamefont
  {{He}}}]{li2022we}%
  \BibitemOpen
  \bibfield  {author} {\bibinfo {author} {\bibfnamefont {J.}~\bibnamefont
  {{Li}}}, \bibinfo {author} {\bibfnamefont {T.}~\bibnamefont {{Guo}}},
  \bibinfo {author} {\bibfnamefont {J.}~\bibnamefont {{Zhao}}}, \ and\ \bibinfo
  {author} {\bibfnamefont {L.}~\bibnamefont {{He}}},\ }\href {\doibase
  10.1103/PhysRevD.106.083021} {\bibfield  {journal} {\bibinfo  {journal}
  {\prd}\ }\textbf {\bibinfo {volume} {106}},\ \bibinfo {eid} {083021}
  (\bibinfo {year} {2022})}\BibitemShut {NoStop}%
\bibitem [{\citenamefont {Kim}(2014)}]{kim2014physics}%
  \BibitemOpen
  \bibfield  {author} {\bibinfo {author} {\bibfnamefont {H.-C.}\ \bibnamefont
  {Kim}},\ }\href {\doibase 10.1103/PhysRevD.89.064001} {\bibfield  {journal}
  {\bibinfo  {journal} {\prd}\ }\textbf {\bibinfo {volume} {89}},\ \bibinfo
  {pages} {064001} (\bibinfo {year} {2014})}\BibitemShut {NoStop}%
\bibitem [{\citenamefont {Wojnar}(2023)}]{Wojnar:2022dvo}%
  \BibitemOpen
  \bibfield  {author} {\bibinfo {author} {\bibfnamefont {A.}~\bibnamefont
  {Wojnar}},\ }\href@noop {} {\bibfield  {journal} {\bibinfo  {journal} {\prd}\
  }\textbf {\bibinfo {volume} {(in press)}} (\bibinfo {year} {2023})},\ \Eprint
  {http://arxiv.org/abs/2208.04023} {arXiv:2208.04023 [gr-qc]} \BibitemShut
  {NoStop}%
\bibitem [{\citenamefont {Sakstein}(2015{\natexlab{a}})}]{sakstein2015testing}%
  \BibitemOpen
  \bibfield  {author} {\bibinfo {author} {\bibfnamefont {J.}~\bibnamefont
  {Sakstein}},\ }\href {\doibase 10.1103/PhysRevD.92.124045} {\bibfield
  {journal} {\bibinfo  {journal} {\prd}\ }\textbf {\bibinfo {volume} {92}},\
  \bibinfo {pages} {124045} (\bibinfo {year} {2015}{\natexlab{a}})}\BibitemShut
  {NoStop}%
\bibitem [{\citenamefont
  {Sakstein}(2015{\natexlab{b}})}]{sakstein2015hydrogen}%
  \BibitemOpen
  \bibfield  {author} {\bibinfo {author} {\bibfnamefont {J.}~\bibnamefont
  {Sakstein}},\ }\href {\doibase 10.1103/PhysRevLett.115.201101} {\bibfield
  {journal} {\bibinfo  {journal} {\prl}\ }\textbf {\bibinfo {volume} {115}},\
  \bibinfo {pages} {201101} (\bibinfo {year} {2015}{\natexlab{b}})}\BibitemShut
  {NoStop}%
\bibitem [{\citenamefont {Olmo}\ \emph {et~al.}(2019)\citenamefont {Olmo},
  \citenamefont {Rubiera-Garcia},\ and\ \citenamefont {Wojnar}}]{Olmo:2019qsj}%
  \BibitemOpen
  \bibfield  {author} {\bibinfo {author} {\bibfnamefont {G.~J.}\ \bibnamefont
  {Olmo}}, \bibinfo {author} {\bibfnamefont {D.}~\bibnamefont
  {Rubiera-Garcia}}, \ and\ \bibinfo {author} {\bibfnamefont {A.}~\bibnamefont
  {Wojnar}},\ }\href {\doibase 10.1103/PhysRevD.100.044020} {\bibfield
  {journal} {\bibinfo  {journal} {Phys. Rev. D}\ }\textbf {\bibinfo {volume}
  {100}},\ \bibinfo {pages} {044020} (\bibinfo {year} {2019})}\BibitemShut
  {NoStop}%
\bibitem [{\citenamefont {Rosyadi}\ \emph {et~al.}(2019)\citenamefont {Rosyadi}
  \emph {et~al.}}]{rosyadi2019brown}%
  \BibitemOpen
  \bibfield  {author} {\bibinfo {author} {\bibfnamefont {A.}~\bibnamefont
  {Rosyadi}} \emph {et~al.},\ }\href {\doibase 10.1140/epjc/s10052-019-7560-3}
  {\bibfield  {journal} {\bibinfo  {journal} {Eur. Phys. J. C}\ }\textbf
  {\bibinfo {volume} {79}},\ \bibinfo {pages} {1} (\bibinfo {year}
  {2019})}\BibitemShut {NoStop}%
\bibitem [{\citenamefont {Lecca}(2021)}]{lecca2021effects}%
  \BibitemOpen
  \bibfield  {author} {\bibinfo {author} {\bibfnamefont {P.}~\bibnamefont
  {Lecca}},\ }in\ \href {\doibase 10.1088/1742-6596/2090/1/012034} {\emph
  {\bibinfo {booktitle} {Journal of Physics: Conference Series}}},\ Vol.\
  \bibinfo {volume} {2090}\ (\bibinfo  {publisher} {IOP Publishing},\ \bibinfo
  {year} {2021})\ p.\ \bibinfo {pages} {012034}\BibitemShut {NoStop}%
\bibitem [{\citenamefont {Delhom-Latorre}\ \emph {et~al.}(2018)\citenamefont
  {Delhom-Latorre}, \citenamefont {Olmo},\ and\ \citenamefont
  {Ronco}}]{delhom2018observable}%
  \BibitemOpen
  \bibfield  {author} {\bibinfo {author} {\bibfnamefont {A.}~\bibnamefont
  {Delhom-Latorre}}, \bibinfo {author} {\bibfnamefont {G.~J.}\ \bibnamefont
  {Olmo}}, \ and\ \bibinfo {author} {\bibfnamefont {M.}~\bibnamefont {Ronco}},\
  }\href {\doibase 10.1016/j.physletb.2018.03.002} {\bibfield  {journal}
  {\bibinfo  {journal} {Phys. Lett. B}\ }\textbf {\bibinfo {volume} {780}},\
  \bibinfo {pages} {294} (\bibinfo {year} {2018})}\BibitemShut {NoStop}%
\bibitem [{\citenamefont {Mestel}\ and\ \citenamefont
  {Ruderman}(1967)}]{mestel1967energy}%
  \BibitemOpen
  \bibfield  {author} {\bibinfo {author} {\bibfnamefont {L.}~\bibnamefont
  {Mestel}}\ and\ \bibinfo {author} {\bibfnamefont {M.}~\bibnamefont
  {Ruderman}},\ }\href {\doibase 10.1093/mnras/136.1.27} {\bibfield  {journal}
  {\bibinfo  {journal} {\mnras}\ }\textbf {\bibinfo {volume} {136}},\ \bibinfo
  {pages} {27} (\bibinfo {year} {1967})}\BibitemShut {NoStop}%
\bibitem [{\citenamefont {Van~Horn}(1968)}]{van1968crystallization}%
  \BibitemOpen
  \bibfield  {author} {\bibinfo {author} {\bibfnamefont {H.}~\bibnamefont
  {Van~Horn}},\ }\href {\doibase 10.1086/149432} {\bibfield  {journal}
  {\bibinfo  {journal} {\apj}\ }\textbf {\bibinfo {volume} {151}},\ \bibinfo
  {pages} {227} (\bibinfo {year} {1968})}\BibitemShut {NoStop}%
\bibitem [{\citenamefont {{De Felice}}\ and\ \citenamefont
  {{Tsujikawa}}(2010)}]{2010LRR....13....3D}%
  \BibitemOpen
  \bibfield  {author} {\bibinfo {author} {\bibfnamefont {A.}~\bibnamefont {{De
  Felice}}}\ and\ \bibinfo {author} {\bibfnamefont {S.}~\bibnamefont
  {{Tsujikawa}}},\ }\href {\doibase 10.12942/lrr-2010-3} {\bibfield  {journal}
  {\bibinfo  {journal} {Living Reviews in Relativity}\ }\textbf {\bibinfo
  {volume} {13}},\ \bibinfo {eid} {3} (\bibinfo {year} {2010})}\BibitemShut
  {NoStop}%
\bibitem [{\citenamefont {{Toniato}}\ \emph {et~al.}(2020)\citenamefont
  {{Toniato}}, \citenamefont {{Rodrigues}},\ and\ \citenamefont
  {{Wojnar}}}]{toniato2020palatini}%
  \BibitemOpen
  \bibfield  {author} {\bibinfo {author} {\bibfnamefont {J.~D.}\ \bibnamefont
  {{Toniato}}}, \bibinfo {author} {\bibfnamefont {D.~C.}\ \bibnamefont
  {{Rodrigues}}}, \ and\ \bibinfo {author} {\bibfnamefont {A.}~\bibnamefont
  {{Wojnar}}},\ }\href {\doibase 10.1103/PhysRevD.101.064050} {\bibfield
  {journal} {\bibinfo  {journal} {\prd}\ }\textbf {\bibinfo {volume} {101}},\
  \bibinfo {eid} {064050} (\bibinfo {year} {2020})}\BibitemShut {NoStop}%
\bibitem [{\citenamefont {Liu}\ and\ \citenamefont
  {L{\"u}}(2019)}]{liu2019properties}%
  \BibitemOpen
  \bibfield  {author} {\bibinfo {author} {\bibfnamefont {H.}~\bibnamefont
  {Liu}}\ and\ \bibinfo {author} {\bibfnamefont {G.}~\bibnamefont {L{\"u}}},\
  }\href {\doibase 10.1088/1475-7516/2019/02/040} {\bibfield  {journal}
  {\bibinfo  {journal} {\jcap}\ }\textbf {\bibinfo {volume} {2019}},\ \bibinfo
  {pages} {040} (\bibinfo {year} {2019})}\BibitemShut {NoStop}%
\bibitem [{\citenamefont {{N{\"a}f}}\ and\ \citenamefont
  {{Jetzer}}(2010)}]{2010PhRvD..81j4003N}%
  \BibitemOpen
  \bibfield  {author} {\bibinfo {author} {\bibfnamefont {J.}~\bibnamefont
  {{N{\"a}f}}}\ and\ \bibinfo {author} {\bibfnamefont {P.}~\bibnamefont
  {{Jetzer}}},\ }\href {\doibase 10.1103/PhysRevD.81.104003} {\bibfield
  {journal} {\bibinfo  {journal} {Phys. Rev. D}\ }\textbf {\bibinfo {volume}
  {81}},\ \bibinfo {eid} {104003} (\bibinfo {year} {2010})}\BibitemShut
  {NoStop}%
\bibitem [{\citenamefont {Olmo}(2005)}]{olmo2005gravity}%
  \BibitemOpen
  \bibfield  {author} {\bibinfo {author} {\bibfnamefont {G.~J.}\ \bibnamefont
  {Olmo}},\ }\href {\doibase 10.1103/PhysRevLett.95.261102} {\bibfield
  {journal} {\bibinfo  {journal} {\prl}\ }\textbf {\bibinfo {volume} {95}},\
  \bibinfo {pages} {261102} (\bibinfo {year} {2005})}\BibitemShut {NoStop}%
\bibitem [{\citenamefont {Avelino}(2012)}]{avelino2012eddington}%
  \BibitemOpen
  \bibfield  {author} {\bibinfo {author} {\bibfnamefont {P.}~\bibnamefont
  {Avelino}},\ }\href {\doibase 10.1088/1475-7516/2012/11/022} {\bibfield
  {journal} {\bibinfo  {journal} {\jcap}\ }\textbf {\bibinfo {volume} {2012}},\
  \bibinfo {pages} {022} (\bibinfo {year} {2012})}\BibitemShut {NoStop}%
\bibitem [{\citenamefont {Jim{\'e}nez}\ \emph {et~al.}(2018)\citenamefont
  {Jim{\'e}nez} \emph {et~al.}}]{jimenez2018born}%
  \BibitemOpen
  \bibfield  {author} {\bibinfo {author} {\bibfnamefont {J.~B.}\ \bibnamefont
  {Jim{\'e}nez}} \emph {et~al.},\ }\href {\doibase
  10.1016/j.physrep.2017.11.001} {\bibfield  {journal} {\bibinfo  {journal}
  {Physics Reports}\ }\textbf {\bibinfo {volume} {727}},\ \bibinfo {pages} {1}
  (\bibinfo {year} {2018})}\BibitemShut {NoStop}%
\bibitem [{\citenamefont {Hernandez-Arboleda}\ \emph
  {et~al.}(2022)\citenamefont {Hernandez-Arboleda}, \citenamefont {Rodrigues},\
  and\ \citenamefont {Wojnar}}]{Hernandez-Arboleda:2022rim}%
  \BibitemOpen
  \bibfield  {author} {\bibinfo {author} {\bibfnamefont {A.}~\bibnamefont
  {Hernandez-Arboleda}}, \bibinfo {author} {\bibfnamefont {D.~C.}\ \bibnamefont
  {Rodrigues}}, \ and\ \bibinfo {author} {\bibfnamefont {A.}~\bibnamefont
  {Wojnar}},\ }\href@noop {} {\  (\bibinfo {year} {2022})},\ \Eprint
  {http://arxiv.org/abs/2204.03762} {arXiv:2204.03762 [astro-ph.GA]}
  \BibitemShut {NoStop}%
\bibitem [{\citenamefont {Kalita}\ \emph {et~al.}(2022)\citenamefont {Kalita},
  \citenamefont {Sarmah},\ and\ \citenamefont {Wojnar}}]{Kalita:2022zki}%
  \BibitemOpen
  \bibfield  {author} {\bibinfo {author} {\bibfnamefont {S.}~\bibnamefont
  {Kalita}}, \bibinfo {author} {\bibfnamefont {L.}~\bibnamefont {Sarmah}}, \
  and\ \bibinfo {author} {\bibfnamefont {A.}~\bibnamefont {Wojnar}},\ }\href
  {\doibase 10.3390/universe8120647} {\bibfield  {journal} {\bibinfo  {journal}
  {Universe}\ }\textbf {\bibinfo {volume} {8}},\ \bibinfo {pages} {647}
  (\bibinfo {year} {2022})}\BibitemShut {NoStop}%
\bibitem [{\citenamefont {Koester}(1972)}]{koester1972outer}%
  \BibitemOpen
  \bibfield  {author} {\bibinfo {author} {\bibfnamefont {D.}~\bibnamefont
  {Koester}},\ }\href@noop {} {\bibfield  {journal} {\bibinfo  {journal}
  {Astron.~Astrophys.}\ }\textbf {\bibinfo {volume} {16}},\ \bibinfo {pages}
  {459} (\bibinfo {year} {1972})}\BibitemShut {NoStop}%
\bibitem [{\citenamefont {Brush}\ \emph {et~al.}(1966)\citenamefont {Brush},
  \citenamefont {Sahlin},\ and\ \citenamefont {Teller}}]{brush1966monte}%
  \BibitemOpen
  \bibfield  {author} {\bibinfo {author} {\bibfnamefont {S.}~\bibnamefont
  {Brush}}, \bibinfo {author} {\bibfnamefont {H.}~\bibnamefont {Sahlin}}, \
  and\ \bibinfo {author} {\bibfnamefont {E.}~\bibnamefont {Teller}},\ }\href
  {\doibase 10.1063/1.1727895} {\bibfield  {journal} {\bibinfo  {journal} {The
  Journal of Chemical Physics}\ }\textbf {\bibinfo {volume} {45}},\ \bibinfo
  {pages} {2102} (\bibinfo {year} {1966})}\BibitemShut {NoStop}%
\bibitem [{\citenamefont {Masuda}\ \emph {et~al.}(2019)\citenamefont {Masuda}
  \emph {et~al.}}]{masuda2019self}%
  \BibitemOpen
  \bibfield  {author} {\bibinfo {author} {\bibfnamefont {K.}~\bibnamefont
  {Masuda}} \emph {et~al.},\ }\href {\doibase 10.3847/2041-8213/ab321b}
  {\bibfield  {journal} {\bibinfo  {journal} {\apjl}\ }\textbf {\bibinfo
  {volume} {881}},\ \bibinfo {pages} {L3} (\bibinfo {year} {2019})}\BibitemShut
  {NoStop}%
\bibitem [{\citenamefont {Laughlin}\ \emph {et~al.}(1997)\citenamefont
  {Laughlin}, \citenamefont {Bodenheimer},\ and\ \citenamefont
  {Adams}}]{laughlin1997end}%
  \BibitemOpen
  \bibfield  {author} {\bibinfo {author} {\bibfnamefont {G.}~\bibnamefont
  {Laughlin}}, \bibinfo {author} {\bibfnamefont {P.}~\bibnamefont
  {Bodenheimer}}, \ and\ \bibinfo {author} {\bibfnamefont {F.~C.}\ \bibnamefont
  {Adams}},\ }\href {\doibase 10.1086/304125} {\bibfield  {journal} {\bibinfo
  {journal} {\apj}\ }\textbf {\bibinfo {volume} {482}},\ \bibinfo {pages} {420}
  (\bibinfo {year} {1997})}\BibitemShut {NoStop}%
\bibitem [{\citenamefont {Cohen}\ \emph {et~al.}(1969)\citenamefont {Cohen},
  \citenamefont {Lapidus},\ and\ \citenamefont {Cameron}}]{cohen1969treatment}%
  \BibitemOpen
  \bibfield  {author} {\bibinfo {author} {\bibfnamefont {J.~M.}\ \bibnamefont
  {Cohen}}, \bibinfo {author} {\bibfnamefont {A.}~\bibnamefont {Lapidus}}, \
  and\ \bibinfo {author} {\bibfnamefont {A.}~\bibnamefont {Cameron}},\ }\href
  {\doibase 10.1007/BF00653943} {\bibfield  {journal} {\bibinfo  {journal}
  {Astrophysics and Space Science}\ }\textbf {\bibinfo {volume} {5}},\ \bibinfo
  {pages} {113} (\bibinfo {year} {1969})}\BibitemShut {NoStop}%
\bibitem [{\citenamefont {Perot}\ and\ \citenamefont
  {Chamel}(2022)}]{perot2022tidal}%
  \BibitemOpen
  \bibfield  {author} {\bibinfo {author} {\bibfnamefont {L.}~\bibnamefont
  {Perot}}\ and\ \bibinfo {author} {\bibfnamefont {N.}~\bibnamefont {Chamel}},\
  }\href {\doibase 10.1103/PhysRevD.106.023012} {\bibfield  {journal} {\bibinfo
   {journal} {\prd}\ }\textbf {\bibinfo {volume} {106}},\ \bibinfo {pages}
  {023012} (\bibinfo {year} {2022})}\BibitemShut {NoStop}%
\bibitem [{\citenamefont {Boshkayev}\ \emph {et~al.}(2012)\citenamefont
  {Boshkayev} \emph {et~al.}}]{boshkayev2012general}%
  \BibitemOpen
  \bibfield  {author} {\bibinfo {author} {\bibfnamefont {K.}~\bibnamefont
  {Boshkayev}} \emph {et~al.},\ }\href {\doibase 10.1088/0004-637X/762/2/117}
  {\bibfield  {journal} {\bibinfo  {journal} {\apj}\ }\textbf {\bibinfo
  {volume} {762}},\ \bibinfo {pages} {117} (\bibinfo {year}
  {2012})}\BibitemShut {NoStop}%
\bibitem [{\citenamefont {Ferrario}\ \emph {et~al.}(2020)\citenamefont
  {Ferrario}, \citenamefont {Wickramasinghe},\ and\ \citenamefont
  {Kawka}}]{ferrario2020magnetic}%
  \BibitemOpen
  \bibfield  {author} {\bibinfo {author} {\bibfnamefont {L.}~\bibnamefont
  {Ferrario}}, \bibinfo {author} {\bibfnamefont {D.}~\bibnamefont
  {Wickramasinghe}}, \ and\ \bibinfo {author} {\bibfnamefont {A.}~\bibnamefont
  {Kawka}},\ }\href {\doibase 10.1016/j.asr.2019.11.012} {\bibfield  {journal}
  {\bibinfo  {journal} {Advances in Space Research}\ }\textbf {\bibinfo
  {volume} {66}},\ \bibinfo {pages} {1025} (\bibinfo {year}
  {2020})}\BibitemShut {NoStop}%
\bibitem [{\citenamefont {Mukhopadhyay}\ \emph {et~al.}(2021)\citenamefont
  {Mukhopadhyay}, \citenamefont {Sarkar},\ and\ \citenamefont
  {Tout}}]{mukhopadhyay2021modified}%
  \BibitemOpen
  \bibfield  {author} {\bibinfo {author} {\bibfnamefont {B.}~\bibnamefont
  {Mukhopadhyay}}, \bibinfo {author} {\bibfnamefont {A.}~\bibnamefont
  {Sarkar}}, \ and\ \bibinfo {author} {\bibfnamefont {C.~A.}\ \bibnamefont
  {Tout}},\ }\href {\doibase 10.1093/mnras/staa3136} {\bibfield  {journal}
  {\bibinfo  {journal} {\mnras}\ }\textbf {\bibinfo {volume} {500}},\ \bibinfo
  {pages} {763} (\bibinfo {year} {2021})}\BibitemShut {NoStop}%
\bibitem [{\citenamefont {Rotondo}\ \emph {et~al.}(2011)\citenamefont {Rotondo}
  \emph {et~al.}}]{rotondo2011relativistic}%
  \BibitemOpen
  \bibfield  {author} {\bibinfo {author} {\bibfnamefont {M.}~\bibnamefont
  {Rotondo}} \emph {et~al.},\ }\href {\doibase 10.1103/PhysRevD.84.084007}
  {\bibfield  {journal} {\bibinfo  {journal} {\prd}\ }\textbf {\bibinfo
  {volume} {84}},\ \bibinfo {pages} {084007} (\bibinfo {year}
  {2011})}\BibitemShut {NoStop}%
\bibitem [{\citenamefont {{de Carvalho}}\ \emph {et~al.}(2014)\citenamefont
  {{de Carvalho}} \emph {et~al.}}]{2014PhRvC..89a5801D}%
  \BibitemOpen
  \bibfield  {author} {\bibinfo {author} {\bibfnamefont {S.~M.}\ \bibnamefont
  {{de Carvalho}}} \emph {et~al.},\ }\href {\doibase
  10.1103/PhysRevC.89.015801} {\bibfield  {journal} {\bibinfo  {journal}
  {\prc}\ }\textbf {\bibinfo {volume} {89}},\ \bibinfo {eid} {015801} (\bibinfo
  {year} {2014})}\BibitemShut {NoStop}%
\bibitem [{\citenamefont {Merkel}\ \emph {et~al.}(2021)\citenamefont {Merkel}
  \emph {et~al.}}]{merkel2021femtosecond}%
  \BibitemOpen
  \bibfield  {author} {\bibinfo {author} {\bibfnamefont {S.}~\bibnamefont
  {Merkel}} \emph {et~al.},\ }\href {\doibase 10.1103/PhysRevLett.127.205501}
  {\bibfield  {journal} {\bibinfo  {journal} {\prl}\ }\textbf {\bibinfo
  {volume} {127}},\ \bibinfo {pages} {205501} (\bibinfo {year}
  {2021})}\BibitemShut {NoStop}%
\bibitem [{\citenamefont {Wojnar}(2021{\natexlab{b}})}]{Wojnar:2021xbr}%
  \BibitemOpen
  \bibfield  {author} {\bibinfo {author} {\bibfnamefont {A.}~\bibnamefont
  {Wojnar}},\ }\href {\doibase 10.1103/PhysRevD.104.104058} {\bibfield
  {journal} {\bibinfo  {journal} {Phys. Rev. D}\ }\textbf {\bibinfo {volume}
  {104}},\ \bibinfo {pages} {104058} (\bibinfo {year}
  {2021}{\natexlab{b}})}\BibitemShut {NoStop}%
\bibitem [{\citenamefont {Dziewonski}\ and\ \citenamefont
  {Anderson}(1981)}]{dziewonski1981preliminary}%
  \BibitemOpen
  \bibfield  {author} {\bibinfo {author} {\bibfnamefont {A.~M.}\ \bibnamefont
  {Dziewonski}}\ and\ \bibinfo {author} {\bibfnamefont {D.~L.}\ \bibnamefont
  {Anderson}},\ }\href {\doibase 10.1016/0031-9201(81)90046-7} {\bibfield
  {journal} {\bibinfo  {journal} {Physics of the earth and planetary
  interiors}\ }\textbf {\bibinfo {volume} {25}},\ \bibinfo {pages} {297}
  (\bibinfo {year} {1981})}\BibitemShut {NoStop}%
\bibitem [{\citenamefont {Donini}\ \emph {et~al.}(2019)\citenamefont {Donini},
  \citenamefont {Palomares-Ruiz},\ and\ \citenamefont
  {Salvado}}]{donini2019neutrino}%
  \BibitemOpen
  \bibfield  {author} {\bibinfo {author} {\bibfnamefont {A.}~\bibnamefont
  {Donini}}, \bibinfo {author} {\bibfnamefont {S.}~\bibnamefont
  {Palomares-Ruiz}}, \ and\ \bibinfo {author} {\bibfnamefont {J.}~\bibnamefont
  {Salvado}},\ }\href {\doibase 10.1038/s41567-018-0319-1} {\bibfield
  {journal} {\bibinfo  {journal} {Nature Physics}\ }\textbf {\bibinfo {volume}
  {15}},\ \bibinfo {pages} {37} (\bibinfo {year} {2019})}\BibitemShut {NoStop}%
\bibitem [{\citenamefont {Biesiada}\ and\ \citenamefont
  {Malec}(2004)}]{biesiada2004new}%
  \BibitemOpen
  \bibfield  {author} {\bibinfo {author} {\bibfnamefont {M.}~\bibnamefont
  {Biesiada}}\ and\ \bibinfo {author} {\bibfnamefont {B.}~\bibnamefont
  {Malec}},\ }\href {\doibase 10.1111/j.1365-2966.2004.07677.x} {\bibfield
  {journal} {\bibinfo  {journal} {\mnras}\ }\textbf {\bibinfo {volume} {350}},\
  \bibinfo {pages} {644} (\bibinfo {year} {2004})}\BibitemShut {NoStop}%
\bibitem [{\citenamefont {Saltas}\ \emph {et~al.}(2018)\citenamefont {Saltas},
  \citenamefont {Sawicki},\ and\ \citenamefont {Lopes}}]{Saltas2018white}%
  \BibitemOpen
  \bibfield  {author} {\bibinfo {author} {\bibfnamefont {I.~D.}\ \bibnamefont
  {Saltas}}, \bibinfo {author} {\bibfnamefont {I.}~\bibnamefont {Sawicki}}, \
  and\ \bibinfo {author} {\bibfnamefont {I.}~\bibnamefont {Lopes}},\ }\href
  {\doibase 10.1088/1475-7516/2018/05/028} {\bibfield  {journal} {\bibinfo
  {journal} {\jcap}\ }\textbf {\bibinfo {volume} {2018}},\ \bibinfo {pages}
  {028} (\bibinfo {year} {2018})}\BibitemShut {NoStop}%
\end{thebibliography}%

\end{document}